\documentclass[manuscript]{aastex61}
\pdfoutput=1

\usepackage{soul,xcolor}
\usepackage{amsmath}

\def\JB{{\rm Jy~beam^{-1}}}
\def\mJB{{\rm mJy~beam^{-1}}}

\def\kms{{\rm km~s^{-1}}}

\received{}
\revised{}
\accepted{}
\submitjournal{ApJ}

\shorttitle{NGC1333 IRAS 4A in Perseus}
\shortauthors{Chuang et al.}

\begin{document}

\title{ALMA Observations toward the S-shaped Outflow and the Envelope around NGC1333 IRAS 4A2}

\correspondingauthor{Chen-Yu Chuang}
\email{cychuang@asiaa.sinica.edu.tw}

\author[0000-0003-2069-9413]{Chen-Yu Chuang}
\affil{Academia Sinica Institute of Astronomy and Astrophysics P.O. Box 23-141, Taipei 106, Taiwan}

\author[0000-0002-8238-7709]{Yusuke Aso}
\affil{Korea Astronomy and Space Science Institute (KASI), 776 Daedeokdae-ro, Yuseong-gu, Daejeon 34055, Republic of Korea}

\author[0000-0001-9304-7884]{Naomi Hirano}
\affil{Academia Sinica Institute of Astronomy and Astrophysics P.O. Box 23-141, Taipei 106, Taiwan}

\author[0000-0002-4317-767X]{Shingo Hirano}
\affil{Department of Earth and Planetary Sciences, Faculty of Sciences Kyushu University, Fukuoka 812-8581, Japan}

\author[0000-0002-0963-0872]{Masahiro N. Machida}
\affil{Department of Earth and Planetary Sciences, Faculty of Sciences Kyushu University, Fukuoka 812-8581, Japan}

\begin{abstract}
We have analyzed the ALMA archival data of the SO ($J_N=6_5-5_4$  and $J_N=7_6-6_5$), CO ($J=2-1$), and CCH ($N=3-2, J=7/2-5/2, F=4-3$) lines from the class 0 protobinary system, NGC1333 IRAS 4A. The images of SO ($J_N = 6_5-5_4$) and CO ($J=2-1$) successfully separate two northern outflow lobes connected to each protostar, IRAS 4A1 and IRAS 4A2. The outflow from IRAS 4A2 shows an S-shaped morphology, consisting of a flattened envelope around IRAS 4A2 with two outflow lobes connected to both edges of the envelope. The flattened envelope surrounding IRAS 4A2 has an opposite velocity gradient to that of the circumbinary envelope. The observed features are reproduced by the magnetohydrodynamic simulation of the collapsing core whose magnetic field direction is misaligned to the rotational axis. Our simulation shows that the intensity of the outfl
ow lobes is enhanced on one side, resulting in the formation of S-shaped morphology. The S-shaped outflow can also be explained by the precessing outflow launched from an unresolved binary with a separation larger than 12 au ($0.04\arcsec$). Additionally, we discovered a previously unknown extremely high velocity component at $\sim$45-90 $\kms$ near IRAS 4A2 with CO. CCH ($J_{N,F}=7/2_{3,4}-5/2_{2,3}$) emission shows two pairs of blobs attaching to the bottom of shell like feature, and the morphology is significantly different from those of SO and CO lines. Toward IRAS 4A2, the S-shaped outflow shown in SO is overlapped with the edges of CCH shells, while CCH shells have the velocity gradients opposite to the flattened structure around IRAS 4A2.

\end{abstract}


\section{INTRODUCTION} 
\label{sec:intro}
In the classical star formation scenario, the magnetic field is assumed to be aligned with the initial rotation axis of the core for simplicity \citep[e.g.][]{nakamura1995,tomisaka1996,machida2008,matsumoto2004,machida2012,machida2013ii}. Thus, in such setting, the outflow propagation direction and the disk rotation axis become parallel to the initial magnetic field. The magnetohydrodynamic (MHD) simulations based on this scenario provides results which the disk rotation axis is parallel to the outflow direction \citep[e.g.,][]{machida11, Banerjee2006}.

However, many observations of dust polarization unveiled the directions of the magnetic fields which are not always aligned with the disk rotation axes and outflow directions. The magnetic field directions in the protostellar envelope scale are randomly aligned with the outflow directions in class 0/I sources \citep[][]{Hull2013}. Similar results are also found in T-Tauri stars \citep[e.g.][]{Menard2004}. These observations imply that the stars are formed in the molecular cloud cores whose initial rotation axes are rondomly aligned with the magnetic fields.

Recently, 3-dimensional MHD simulations were conducted to reveal the evolution of the protostellar core whose initial rotation axis and outflow direction are misaligned with magnetic field \citep[e.g.][]{matsumoto2004,Joos2012, Li2013}. It is, therefore, important to study the detailed structure and kinematics of the outflow/envelope systems with misaligned magnetic fields, and compare them with the MHD simulations.

We choose NGC1333 IRAS 4A as our target because this object is a good example showing the misalignment between magnetic field and outflow direction. IRAS 4A is a Class 0 binary system located in star forming region NGC1333, which is 293 $\pm$ 22 pc away from the solar system \citep[][]{Zucker2018}. Previous high resolution observations toward IRAS 4A suggested that this system consists of two compact continuum sources having a separation of $1.8 \arcsec$ \citep[e.g.][]{Looney2000, Reipurth2002} ($\sim 530$ au). The names, IRAS 4A1 and IRAS 4A2, are assigned to the eastern and western continuum peaks, respectively. The two sources are embedded in a circumbinary-envelope traced by the C$^{17}$O emission, which shows a North-West (redshifted) to South-East (blueshifted) velocity gradient \citep[][]{Ching2016}. Interestingly, the ammonia line emission shows an opposite velocity gradient near IRAS 4A2, which is implied as a Keplarian disk rotating oppositely to the circumbinary-envelope \citep[][]{choi07}. 

IRAS 4A is associated with a large (a few arcminutes) scale bipolar outflow, which is traced by several molecular lines such as CO, SiO, etc \citep[e.g.][]{Choi2001,Santangelo2015, yildiz2012}. The interferometric image of SiO 1--0 revealed that the Northern red-shifted lobe is bent to east at $\sim$20$\arcsec$ away from the continuum peaks, while two blue-shifted southern lobes are ejected toward south-east and south-west \citep{Choi2001}.
The magnetic field in the IRAS 4A envelope shows a classical pinched hour-glass shape with an orientation angle of $61 \arcdeg$ \citep[][]{doi2020}, which is misaligned with the rotation axis of the circumbinary-envelope ($38 \arcdeg$). On smaller scale, the magnetic field is misaligned with the axes of the outflows from both IRAS 4A1 (Northern lobe: $19 \arcdeg$, Southern lobe: $-9 \arcdeg$) and IRAS 4A2 ($19 \arcdeg$)\citep[][]{Ching2016}. Furthermore, orientation of the magnetic field is also misaligned with the angular momentum vector of the IRAS 4A2 disk ($19 \arcdeg$) \citep[][]{Choi2011}.

In this paper, we present the ALMA (Atacama Large Millimeter/sub-millimeter Array) observations with higher resolution (0\farcs3--0\farcs7 resolution) toward IRAS 4A system. The molecular line observations include two transitions of SO ($J_N = 6_5-5_4$ and $J_N = 7_6-6_5$), CO ($J = 2-1$) and CCH ($N = 3-2$, $J = 7/2-5/2$). 
We have studied the morphology and kinematics of the outflow and envelope around IRAS 4A2 in detail, and compared the observational results with the recent resistive MHD simulations with misaligned magnetic fields and rotation (Hirano \& Machida 2019).

\section{OBSERVATIONS} \label{sec:observations}
We used three sets of ALMA archival data, ALMA2013.1.01102 in 262 GHz (P.I. N. Sakai), ALMA2017.1.00053 and ALMA2013.1.00031 in 230 GHz (P.I. J. Tobin). 

The 262 GHz observations were carried out on June 13 in 2015. The total observing time and the on-source time were 57 minutes and 26 minutes, respectively. The number of antennas was 35. The minimum projected baseline length was 21.3 m, which corresponds to the maximum recoverable size of $6\farcs 72$ (1969 au). The spectral windows for the SO ($J_N=7_6-6_5$) and CCH ($N=3-2, J=7/2-5/2, F=4-3$) lines have 960 channels covering 58.594 MHz at a frequency resolution of 61.035 kHz. In making maps, 7 channels were binned for both lines. The resultant velocity resolution is $0.49\ \kms$ (427.239 kHz). There is no spectral window assigned to the continuum in this data set. Therefore, we separated the continuum from the line observation.

The 230 GHz observations were carried out on September 27 in 2015 (ALMA2013.1.00031) and from 2017 December to 2018 September (ALMA2017.1.00053). For ALMA2013.1.00031, the total observing time and the on-source time were 107 minutes and 3.5 minutes, respectively. The number of antennas was 33. The minimum projected baseline length was 43.3 m, which corresponds to the maximum recoverable size of $3\farcs91$ (1146 au). For ALMA2017.1.00053, the total observing time and the on-source time were 347.4 minutes and 18.14 minutes, respectively. The number of antennas was 45. The minimum projected baseline length was 15.1 m, which corresponds to the maximum recoverable size of $11\farcs2$ ($\sim 3000$ au). In ALMA2017.1.00053, the spectral windows for the SO ($J_N=6_5-5_4$) and CO $(J=2-1)$ lines have 960 and 1920 channels covering 58.593 MHz and 234.375 MHz, at the frequency resolution of 61.035 kHz and 244.141 kHz, respectively. In ALMA2013.1.00031, one spectral window covering 232.5-234.5 GHz was assigned to continuum emission.

We used Common Astronomical Software Applications (CASA) for image processing. The raw visibility data of ALMA2013.1.00031, ALMA2017.1.00053 and ALMA2013.1.01102 were calibrated by CASA version 4.5.0, 5.4.0, and 4.3.1, respectively. The calibrated visibilities were Fourier transformed and CLEANed by the task {\it tclean} with the natural weighting and a threshold of 1$\sigma$. We also performed self-calibration for the continuum data using tasks in CASA ({\it tclean}, {\it gaincal}, and {\it applycal}). First, the phase was calibrated with the time bin of 3 scans ($\sim 60$s). Then, using the derived gain table, the amplitude and the phase were calibrated together. The self-calibration improved the RMS noise level of the continuum maps by a factor of $\sim 2$. The obtained calibration tables for the continuum data were also applied to the line data. The noise levels of the line maps were measured in emission-free channels. The parameters of our observations mentioned above and others are summarized in Table \ref{ch4:tab:obs}.

\begin{deluxetable*}{ccccccc}
\tabletypesize{\footnotesize}
\tablecaption{Summary of the ALMA observational parameters \label{ch4:tab:obs}}
\tablehead{
\colhead{Data set} & \multicolumn{2}{c}{262GHz} & \multicolumn{4}{c}{230GHz}\\
\colhead{Project Code} & \multicolumn{2}{c}{ALMA2013.1.01102.S} & \multicolumn{2}{c}{ALMA2013.1.00031.S} & \multicolumn{2}{c}{ALMA2017.1.00053.S}\\
\colhead{Date} & \multicolumn{2}{c}{13-Jun-2015} & \multicolumn{2}{c}{27-Sep-2015} & \multicolumn{2}{c}{\begin{tabular}{c}
17-Dec-2017, 07-Jan-2018, 11-Jan-2018,\\ 18-Jan-2018, 20-Sep-2018
\end{tabular}}\\
\colhead{Projected baseline length} & \multicolumn{2}{c}{
\begin{tabular}{c}
21.3 - 783.5 m\\(18.2 - 669.7 k$\lambda$)
\end{tabular}
}
& \multicolumn{2}{c}{
\begin{tabular}{c}
43.3 - 2270 m\\(31.6 - 1669.1 k$\lambda$)
\end{tabular}
} & \multicolumn{2}{c}{
\begin{tabular}{c}
15.1 - 2516.9 m\\(11.6 - 1931 k$\lambda$)
\end{tabular}
}\\
\colhead{Maximum recoverable scale} & \multicolumn{2}{c}{\begin{tabular}{c}
$6\farcs72$\\(2000 au)
\end{tabular}
}
& \multicolumn{2}{c}{
\begin{tabular}{c}
$3\farcs91$\\(1000 au)
\end{tabular}
} & \multicolumn{2}{c}{
\begin{tabular}{c}
$11\farcs2$\\(3000 au)
\end{tabular}
} \\
\colhead{Primary beam} & \multicolumn{2}{c}{$23.6\arcsec$} & \multicolumn{2}{c}{$25.5\arcsec$} & \multicolumn{2}{c}{$25.5\arcsec$}\\
\colhead{Bandpass calibrator} & \multicolumn{2}{c}{J0237$+$2848} &
\multicolumn{2}{c}{J0237$+$2848} &
\multicolumn{2}{c}{J0237$+$2848}\\
\colhead{Flux calibrator} & \multicolumn{2}{c}{Titan} & \multicolumn{2}{c}{Titan} & \multicolumn{2}{c}{J0237$+$2848}\\
\colhead{Phase calibrator} & \multicolumn{2}{c}{J0319$+$4130} &
\multicolumn{2}{c}{J0319$+$4130} &
\multicolumn{2}{c}{J0336$+$3218}\\
\colhead{Phase center coordinate (J2000)} & \multicolumn{2}{c}{$03^{\rm h}29^{\rm m}$10\fs 51, $31^{\circ}13\arcmin 31\farcs 3$} & \multicolumn{2}{c}{$03^{\rm h}29^{\rm m}$10\fs 536, $31^{\circ}13\arcmin 30\farcs 93$} & \multicolumn{2}{c}{$03^{\rm h}29^{\rm m}$10\fs 536, $31^{\circ}13\arcmin 30\farcs 93$}\\}
\startdata
 &
\begin{tabular}{c}
Rest \\frequency\\(GHz)
\end{tabular}
&
\begin{tabular}{c}
Center \\frequency\\(GHz)
\end{tabular}
&
\begin{tabular}{c}
Velocity\\resolution\\($\kms$)
\end{tabular}
&
\begin{tabular}{c}
Total\\bandwidth\\(MHz)
\end{tabular}
&
\begin{tabular}{c}
Beam\\Size\\(P.A.)
\end{tabular}
&
\begin{tabular}{c}
RMS noise level\\($\mJB$)
\end{tabular}\\
\hline
1.3 mm Continuum & - & 232.529000 & - & 2000 & $0\farcs 26\times 0\farcs16\ (25\arcdeg)$ & 0.28 \\
$^{12}$CO ($J=2-1$) & 230.538000 & - & 0.32 & 234.375 & $0\farcs 29\times 0\farcs21\ (-26\arcdeg)$ & 3.4 \\
$^{32}$SO ($J_N=6_5-5_4$) & 219.949433 & - & 0.083 & 58.593 & $0\farcs 29\times 0\farcs21\ (-26\arcdeg)$ & 3.88 \\
SO ($J_N=7_6-6_5$) & 261.843721 & - & 0.49 & 58.594 & $0\farcs 65\times 0\farcs35\ (-28\arcdeg)$ & 2.38 \\
CCH ($N=3-2, J=7/2-5/2$) &&&&&\\
$F=4-3$ & 262.00426 & - & 0.49 & 58.594 & $0\farcs 75\times 0\farcs4\ (-32\arcdeg)$ & 2.11 \\
$F=3-2$ & 262.00648 & - & 0.49 & 58.594 & $0\farcs 75\times 0\farcs4\ (-32\arcdeg)$ & 2.11 \\
\enddata
\end{deluxetable*}
\newpage

\section{RESULTS} \label{sec:results}
\subsection{1.3mm Continuum} \label{sec:cont13} 
\begin{figure}[ht!]
\epsscale{1.2}
\plotone{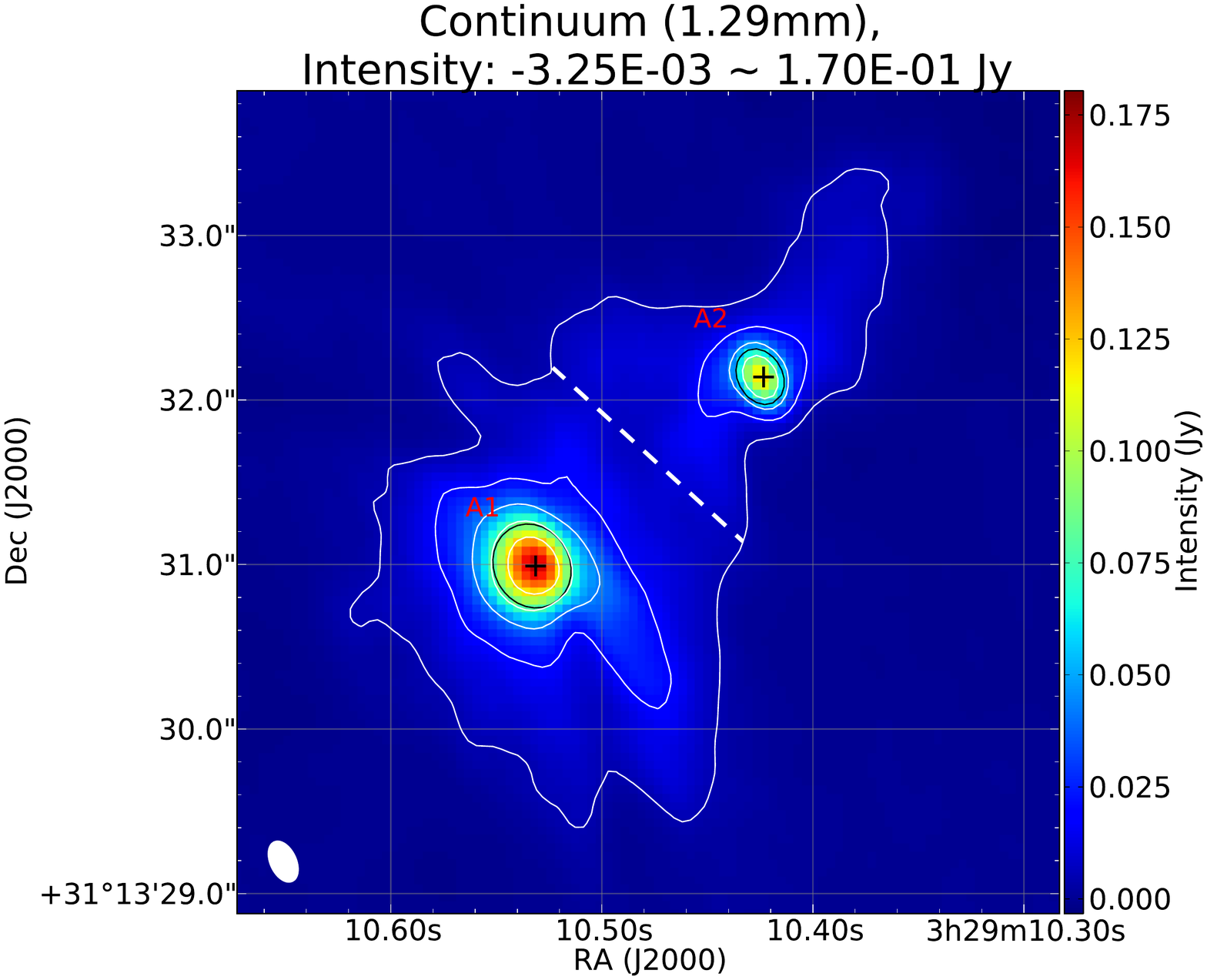}
\caption{The map of 1.3 mm continuum emission of IRAS 4A1 and IRAS 4A2 after primary beam correction. White contour levels are $1,4,9,16,25 \times 3\sigma$, where 1 $\sigma$ corresponds to $0.28\ \mJB$. The color map shows intensity in the unit of $\JB$. Black plus signs and contours denote two peak positions and the area used for 2-component 2D Gaussian fittings, respectively. A white filled ellipse at the bottom-left corner denotes the ALMA synthesized beam; $0\farcs26 \times 0\farcs16,\ {\rm P.A.}=25^{\circ}$. The white dashed line separates the extended component of IRAS 4A1 and IRAS 4A2.
\label{fig:cont13}}
\end{figure}

Figure \ref{fig:cont13} shows a map of 1.3 mm continuum emission from the IRAS 4A system at an angular resolution of $0\farcs26 \times 0\farcs16$. IRAS 4A is clearly resolved into two local peaks, which correspond to IRAS 4A1 and IRAS 4A2. We applied a 2D two-component Gaussian fitting to derive the peaks. The fitting is conducted using the area bounded by the black contours in Figure \ref{fig:cont13}. The derived peak locations in IRAS 4A1 and IRAS 4A2 are $\alpha ({\rm J2000})=03^{\rm h}29^{\rm m}10\fs 533,\ \delta ({\rm J2000})=31\arcdeg 13\arcmin 30\farcs 99$ and $\alpha ({\rm J2000})=03^{\rm h}29^{\rm m}10\fs 425,\ \delta ({\rm J2000})=31\arcdeg 13\arcmin 32\farcs 14$, respectively. The angular separation between IRAS 4A1 and IRAS 4A2 continuum peaks is $1\farcs80$ ($\sim 530\ {\rm au}$), which is consistent with the previous measurements \citep{Sahu2019, Sepulcre2017, su2019}. 

The continuum emission around IRAS 4A1 shows a compact elliptical structure with an extension to the southwest. Its peak intensity and beam-deconvolved size are $165.2\pm3.7\ \mJB$ and $0\farcs45 \times 0\farcs43$ (P.A.=38$\arcdeg$), respectively.
The observed peak intensity corresponds to the brightness temperature of $T_{\rm b}=89$ K, and the Planck temperature of $T_{\rm p}=95$ K.
In order to compare the peak intensity with previous research, we convolved our map to the same beam size ($0\farcs3 \times 0\farcs2$) as that of \citet[][]{Sahu2019}. The peak intensity after convolution is $199.1\ \mJB$, which corresponds to the brightness temperature of $T_{\rm b}=75$ K and the Planck temperature of $T_{\rm p}=80.4$ K. 
These are higher than the peak brightness temperature of $T_{\rm b}=57$ K and Planck temperature of $T_{\rm p}=65.2$ K at 0.84 mm \citep[][]{Sahu2019}. The higher brightness temperature at 1.3 mm indicates that the 1.3 mm continuum having the lower optical depth than 0.84 mm traces the inner region with higher temperature. In addition, the continuum emission from IRAS 4A1 is considered to be optically thick at 1.3 mm. To estimate the dust mass in the extended component (emission above 3$\sigma$ in Figure \ref{fig:cont13}) and compact component (emission within the black contour in Figure \ref{fig:cont13}) of IRAS 4A1 through equation $M_d=\frac{D^2F_\nu}{\kappa_\nu B_\nu(T_d)}$, we assume a gas-to-dust ratio of 100, a dust opacity of 0.01 $cm^{2}\ g^{-1}$ for the compact component and 0.008 $cm^{2}\ g^{-1}$ for the extended component \citep[][]{ossenkopf1994}, a dust temperature of 90 K for the compact component and 60 K for extended component \citep[][]{Sahu2019}. The masses of the extended component and the compact component are estimated to be 0.86 M$_\odot$ (1.51 Jy) and 0.14 M$_\odot$ (0.46 Jy), respectively. The mass of the extended component is a factor of two higher than that of \citet[][]{Sahu2019} probably due to a different aperture size when estimating the total flux and the slightly higher dust mass opacity used for our calculation. 
It should be noted that the estimated mass of the compact component is a lower limit due to the optically thick condition.

The continuum emission from IRAS 4A2 shows a compact elliptical structure smaller than that from IRAS 4A1 and with an extension along the northwest-southeast direction. Its peak intensity and the beam-deconvolved size are $117.3\pm3.2\ \mJB$ and $0\farcs23 \times 0\farcs21$ (P.A.=50$\arcdeg$), respectively. The observed peak intensity corresponds to the brightness temperature of $T_{\rm b}=63\ {\rm K}$, and the Planck temperature of $T_{\rm p}=69.0\ {\rm K}$. After convolving our image to the beam size of $0\farcs3 \times 0\farcs2$, the peak intensity, the corresponding brightness temperature and the Planck temperature are $114.2\ \mJB$, $T_{\rm b}=43\ {\rm K}$ and $T_{\rm p}=48.4\ {\rm K}$, respectively. These are the same as the peak brightness temperature and Planck temperature at 0.84 mm. This indicates that the beam averaged continuum emission from IRAS 4A2 is also optically thick at both 0.84 mm and 1.3 mm. 
The line emission observed toward the continuum peak of IRAS 4A2 at these wavebands \citep{Sahu2019, Sepulcre2017, su2019} suggests that the beam filling factor of the optically thick continuum emission is rather small. Assuming the same parameters as those adopted for IRAS 4A1 except that the temperature of the compact component is 65 K for IRAS 4A2, the gas masses of the extended component and the compact component of IRAS 4A2 are estimated to be 0.32 M$_\odot$ (0.57 Jy) and 0.06 M$_\odot$ (0.13 Jy), respectively. The area of the compact component of IRAS 4A2 continuum is smaller than the disk traced by ammonia \citet[][]{Choi2010}, while the ammonia emitting area corresponds to the 12 $\sigma$ contour in our continuum map.

Additionally, \citet[][]{Santangelo2015} claimed to detect another continuum source, IRAS 4A3, at $4\arcsec$ north of IRAS 4A2 continuum peak at the wavelength of 1.3 mm. However, there is no counterpart of IRAS 4A3 in our continuum image of the same wavelength. 
The upper limit of the continuum flux at the IRAS 4A3 position is $1.2\mJB$ ($3\sigma$, after primary beam correction), which corresponds to $T_B=0.65$ K. This source is not detected by other observations either; \citet[][]{tobin2018ii} attributed this to the artificial effects in PdBI observations, and  \citet[][]{maury2019} to the continuum emission produced by the interaction between outflow and envelope.

\newpage

\subsection{CO ($J = 2 - 1$) outflow}
\label{sec:co}
\begin{figure*}[ht!]
\epsscale{0.95}
\plotone{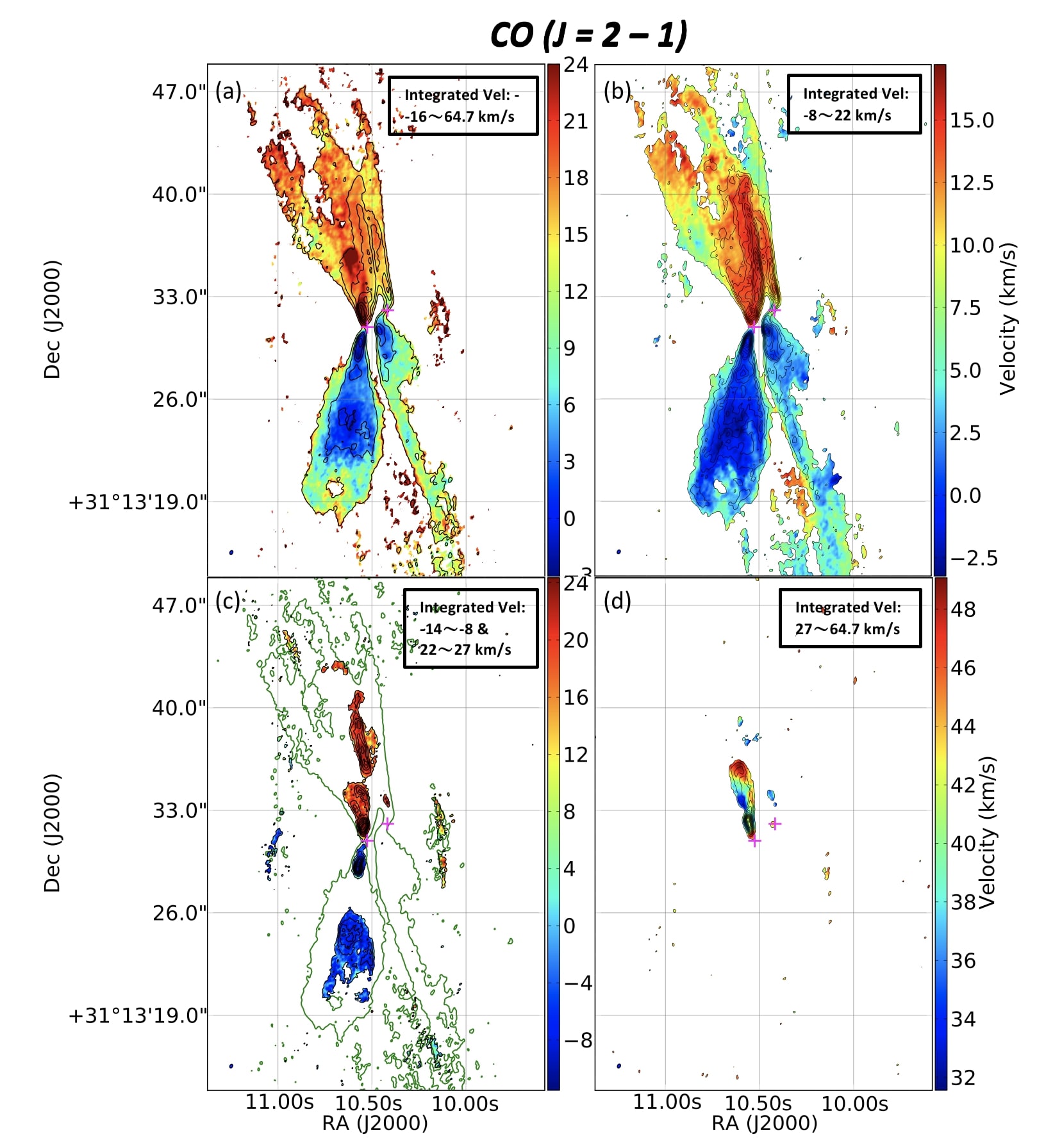}
\caption{Integrated-intensity (moment 0; contours) and mean-velocity (moment 1; color) maps with integrated velocity ranges of (a) $V_{\rm LSR}=-16\sim64.7\ \kms$, (b) $V_{\rm LSR}=-8-22\ \kms$, (c) $V_{\rm LSR}=-14\sim8$ and $22-27\ \kms$, and (d) $V_{\rm LSR}=27 - 64.7\ \kms$. The contour levels are $1,2,3,4,\dots \times 3\sigma$. 1 $\sigma$ corresponds to (a) 43.26, (b) 41.9, (c) 21.9, and (d) $28.6\ \mJB~\kms$. Magenta plus signs denote the continuum peak positions of IRAS 4A1 and IRAS 4A2. Blue-filled ellipses at the bottom-left corners denote the ALMA synthesized beam size: $0\farcs29 \times 0\farcs21,\ {\rm P.A.}=-26^{\circ}$. The green contour in panel (c) denotes the 3 $\sigma$ cutoff in panel (a). We only choose the peaks with counter parts in SO ($J_N=7_6-6_5$) high velocity moment maps (overlapped with SO knots, see Figure \ref{fig:so}d). 
\label{fig:co}}
\end{figure*}

Figure \ref{fig:co} shows the CO ($J=2-1$) moment 0 (integrated-intensity) and moment 1 (mean-velocity) maps with different velocity ranges.
Figure \ref{fig:co}a clearly shows that each protostar, IRAS 4A1 and IRAS 4A2, drives its own outflow toward the north and the south. The two northern lobes are redshifted and bent toward the northeast. The two southern lobes are blueshifted; the one from IRAS 4A1 extends to the southeast, while the one from IRAS 4A2 extends to the southwest. The typical velocity range of the IRAS 4A1 outflow is $\delta V = 0-20\ \kms$ with respect to the systematic velocity of $V_{\rm LSR}=7.35\ \kms$ (see section \ref{sec:pv}). The typical velocity range of the IRAS 4A2 outflow is $\delta V$ $= 0-15\ \kms$, which is smaller than that of the IRAS 4A1 outflow. The outflow from IRAS 4A2 shows an S-shaped morphology with two curved lobes elongated along a P.A.= 20$\arcdeg$ direction. The central $\pm 1\arcsec$ region of the S-shaped feature is elongated along the northwest-southeast direction (P.A. 148\arcdeg). The gap at the center is due to the continuum subtraction effect. \citet[][]{Santangelo2015} reported the S-shape bending of the IRAS 4A2 outflow at $\pm$4\arcsec from the driving source. This bending is also seen in the blueshifted lobe of our CO map. On the other hand, the S-shaped feature seen in our map at the central 2\arcsec region is not clearly resolved in the images of \citet[][]{Santangelo2015}. At a larger scale of $\pm 20\arcsec$ , which is beyond our field of view, the outflow lobes show further bent toward a direction of P.A.= 45$\arcdeg$ \citep[e.g.][]{yildiz2012, choi2005}. Toward the center of IRAS 4A1, CO is observed as absorption because of the opaque continuum. The absorption at IRAS 4A1 is also observed in the SO lines presented in Section \ref{sec:so} and many other lines at 0.85 mm \citep[][]{Sahu2019}.

Figure \ref{fig:co}c shows that the high velocity component of the IRAS 4A1 outflow consists of two red shifted components at the northern lobe and two blue shifted components in the southern lobe. The southernmost part of the blueshifted lobe shows an up-side-down U-shaped feature. The emission at the very high velocity (Figure \ref{fig:co}d, from 20 to 50 $\kms$) is seen only in the northern redshifted lobe of the IRAS 4A1 outflow. The location of this component is slightly east to those of the emission seen in the low (Figure \ref{fig:co}b, from $-$8 to 22 $\kms$) and high (Figure \ref{fig:co}c, from $-$14 to $-$8 and from 22 to 27 $\kms$) velocity ranges, which is consistent with that of the extremely high-velocity peaks (from 41 to 55 $\kms$) in \citet[][]{Santangelo2015}. There is no blueshifted counterpart of this very high velocity emission in the southern lobe.

\begin{figure*}[ht!]
\epsscale{1.0}
\plotone{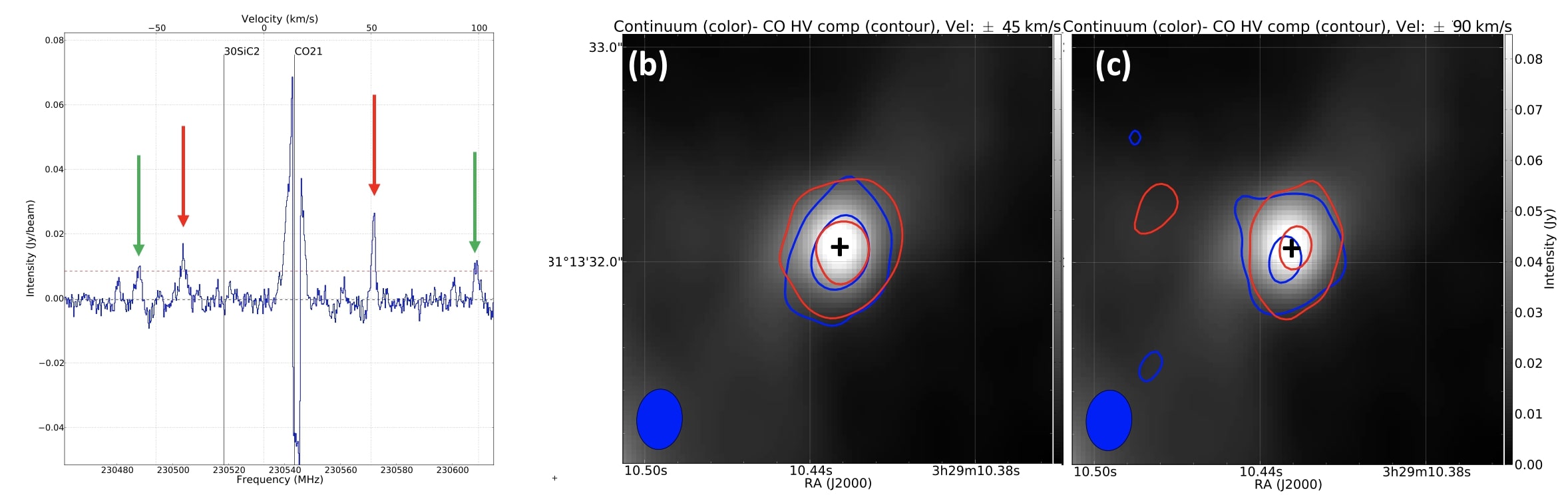}
\caption{(a) The spectral line profile of CO $2-1$. A circle area centered at IRAS 4A2 continuum peak with the radius of $0.25\arcsec$ is used for this line profile. The red dashed line denotes the $3\sigma$ level of the line profile. Two spectral lines, CO $2-1$ and $^{30}$SiC$_2$, are placed for reference, with there rest frequencies offset by $V_{sys} = 7.35 \kms$. The red arrows and green arrows point out the local peaks correspond to panel (b) and (c), respectively. (b) Integrated-intensity (moment 0; contours) and continuum (color) maps with integrated velocity ranges of $V_{LSR}$ = $-46 - -43 \kms$ and $V_{LSR}$ = $42 - 46 \kms$ for blue and red contours, respectively. (c) Same as panel (b), but with integrated velocity ranges of $V_{LSR}$ = $-66 - -63 \kms$ and $V_{LSR}$ = $90 - 93 \kms$ for blue and red contours, respectively. Blue-filled ellipses at the bottom-left corners of panels (b) and (c) denote the ALMA synthesized beam: $0\farcs29 \times 0\farcs21,\ {\rm P.A.}=-26^{\circ}$.
\label{fig:COA2HV}}
\end{figure*}

Although Figure \ref{fig:co}c does not show clearly, the IRAS 4A2 outflow also shows very high velocity components at $\sim \pm 45 \kms$ and $\sim \pm 90 \kms$. 
As shown in Figure \ref{fig:COA2HV}a, the spectral line toward the IRAS 4A2 continuum peak consists of five velocity components above the 3 $\sigma$ level. In addition to the central one that is associated with the low velocity component, there are two peaks at $\sim 45\ \kms$ and $\sim -45\ \kms$ and additional two peaks at $\sim 95\ \kms$ and $\sim -60\ \kms$. These additional four peaks are likely to be the very high velocity components from IRAS 4A2 because there is no spectral line having a frequency corresponds to these emission peaks. The spatial distributions of these extremely high velocity components are shown in Figure \ref{fig:COA2HV}b and Figure \ref{fig:COA2HV}c. The integrated velocity ranges are from $V_{\rm LSR}= -46$ to $-$43 km s$^{-1}$ (blue) and from 42 to 46 km s$^{-1}$ (red) for Figure \ref{fig:COA2HV}b, and from $V_{\rm LSR}= -66$ to $-$63 km s$^{-1}$ (blue) and from 90 to 93 km s$^{-1}$ (red) for Figure \ref{fig:COA2HV}c. The morphology of these compact components is not clearly resolved under the current resolution. 

\clearpage

\subsection{SO outflow}
\label{sec:so}
\begin{figure*}[ht!]
\epsscale{0.83}
\plotone{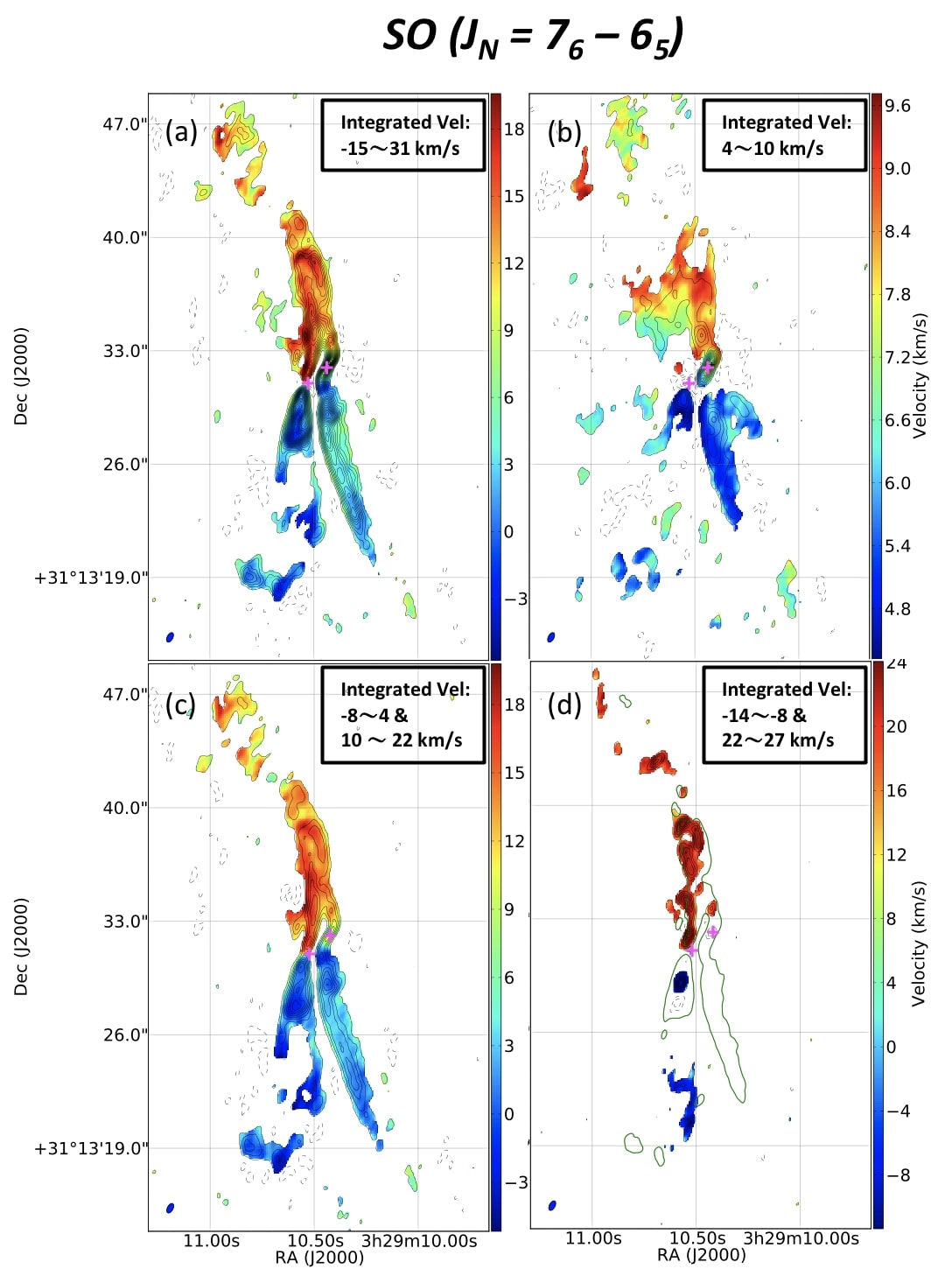}
\caption{Integrated-intensity (moment 0; contours) and mean-velocity (moment 1; color) maps with integrated velocity ranges of (a) $V_{\rm LSR}=-15\sim31 \kms$, (b) $V_{\rm LSR}=4-10\ \kms$, (c) $V_{\rm LSR}=-8\sim4$ and $10-22\ \kms$, and (d) $V_{\rm LSR}=-14\sim8$ and $22-27\ \kms$. The contour levels are $1,4,9,16,\dots \times 3\sigma$ for (a), (b), and (c) while $1,2,3,\dots \times 3\sigma$ for (d). 1 $\sigma$ corresponds to (a) 13.5, (b), 3.9, (c) 17.0, and (d) 6.4 $\mJB~\kms$. Magenta plus signs denote the continuum peak positions of IRAS 4A1 and IRAS 4A2. Blue-filled ellipses at the bottom-left corners denote the ALMA synthesized beam: $0\farcs65 \times 0\farcs35,\ {\rm P.A.}=-28^{\circ}$. The green contour in (d) denotes the 9 $\sigma$ cut off in the moment 0 of panel (a). 
\label{fig:so}}
\end{figure*}

Figure \ref{fig:so} shows SO ($J_N=7_6-6_5$) moment 0 and moment 1 maps with different velocity ranges. As seen in the CO emission, two bipolar outflows driven from two protostars are clearly traced by the SO ($J_N=7_6-6_5$) emission. Overall features of the SO outflows such as outflow directions, typical velocity ranges, and morphology are the similar to those of the CO outflows. 
As shown in Figure \ref{fig:so}a, the SO emission from the northern lobe of the IRAS 4A1 outflow is enhanced at the western wall where the two redshifted lobes from two outflows are overlapping.
The S-shaped feature associated with IRAS 4A2 is also seen in the SO emission.
These outflow features are also seen in the images of another transition of SO ($J_N = 6_5-5_4$) presented in Figure \ref{fig:app_sol} in Appendix \ref{sec:app_sol}.
The spatially extended features from the outflows are better traced by the lower resolution SO ($J_N = 7_6-6_5$) images.
As in the case of CO, both SO lines are also observed as absorption toward IRAS 4A1.

In the very low velocity range (Figure \ref{fig:so}b, from 4 to 10 $\kms$), the SO emission from two outflows shows spatially extended structures; the SO emission in the northern lobe of the IRAS 4A1 outflow extends to the east and fills the lobe. Meanwhile, the S-shaped feature in the IRAS 4A2 outflow is less clear. 
Instead, the outflow morphology is rather bipolar in this velocity range.
Two bright blobs in the blueshifted velocity range labeled SOA2S1 and SOA2S2 in Figure \ref{fig:cchso} delineate the eastern and western walls of the southern lobe.
Although the morphology of the northern lobe is contaminated by the IRAS 4A1 outflow, the redshifted blob (SOA2N) and the emission extending to the north delineate the western wall of the triangle-shaped northern lobe.
The eastern wall of this northern lobe is likely to be the emission ridge extending to the northeast as shown in the dashed line in Figure \ref{fig:cchso}. It should be noted that this very low-velocity component in CO is contaminated by the ambient component, which is spatially extended and resolved out by the interferometer. 

The morphology of the low velocity components (Figure \ref{fig:so}c) is similar to that of the whole outflow (Figure \ref{fig:so}a), whereas that of the high velocity emission (Figure \ref{fig:so}d) is knotty. The peak positions of the SO knots generally align with the CO high velocity components, with a few exceptions at the west of the components. These exceptions are likely to be associated with the northern lobe of the IRAS 4A2 outflow because they are located west to the CO outflow and inline with the overall IRAS 4A2 outflow extension. There is also an additional component at the south of those knots, which partially coincides with the CO emission in the high velocity range. Unfortunately, the SO emission in the very high velocity ranges was not covered by this observation. 

\newpage

\subsection{Column Density and Temperature}
\label{sec:anal}
\begin{figure*}[ht!]
\epsscale{1.0}
\plotone{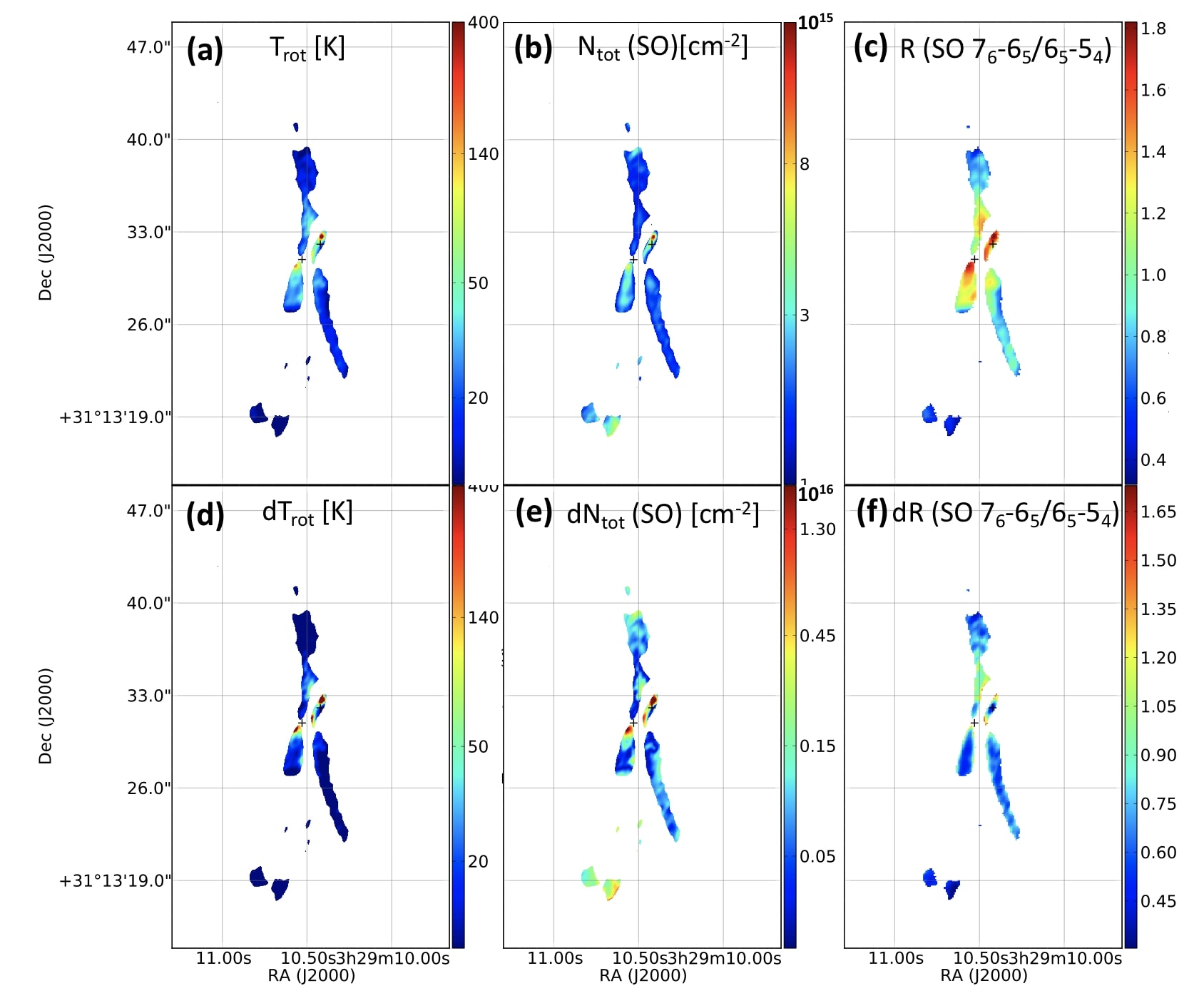}
\caption{(a) The excitation temperature distribution. (b) The column density distribution. (c) The line ratio of SO ($J_N=7_6-6_5$) over SO ($J_N=6_5-5_4$) emission. The uncertainty estimation for the physical quantities in panels (a), (b), and (c) are in panels (d), (e), and (f), respectively. The uncertainties are derived following the law of error propagation.
\label{fig:TNmap}}
\end{figure*}

To further understand the physical properties of the SO outflows, we derive the excitation temperature and column density of SO using the two SO transitions. 
The moment 0 map of SO ($J_N=6_5-5_4$) (Figure \ref{fig:app_sol}a) was convolved to the same beam size as that of SO ($J_N=7_6-6_5$) moment 0 map (Figure \ref{fig:so}. 
Then, we assumed the optically thin and local thermal equilibrium (LTE) assumption, and derived the excitation temperature and the column density of SO following the method of \citet[][]{goldsmith1999}. Figure \ref{fig:TNmap}(a) and Figure \ref{fig:TNmap}(b) show the excitation temperature and column density distributions, respectively. On a larger scale, the excitation temperature and column density are higher at the vicinity of IRAS 4A1 and IRAS 4A2, whereas these are lower in the northern and southern outflow lobes. This trend is consistent with that of SO$_2$ derived from the rotational diagram analysis \citep[][]{taquet2020}. In a smaller scale ($4\arcsec$ around IRAS 4A2), the excitation temperature and column density show the maxima at the northwest of IRAS 4A2 and the northern tip of southern IRAS 4A1 outflow, where the line ratio in Figure \ref{fig:TNmap}(c) shows the enhancement. However, the derived  excitation temperature and column density values in these regions are not reliable because of the large uncertainties shown in (Figure \ref{fig:TNmap}(d) and Figure \ref{fig:TNmap}(e).

\newpage

\subsection{Velocity Structure in the IRAS 4A2 Envelope}
\label{sec:pv}
\begin{figure}[ht!]
\epsscale{0.8}
\plotone{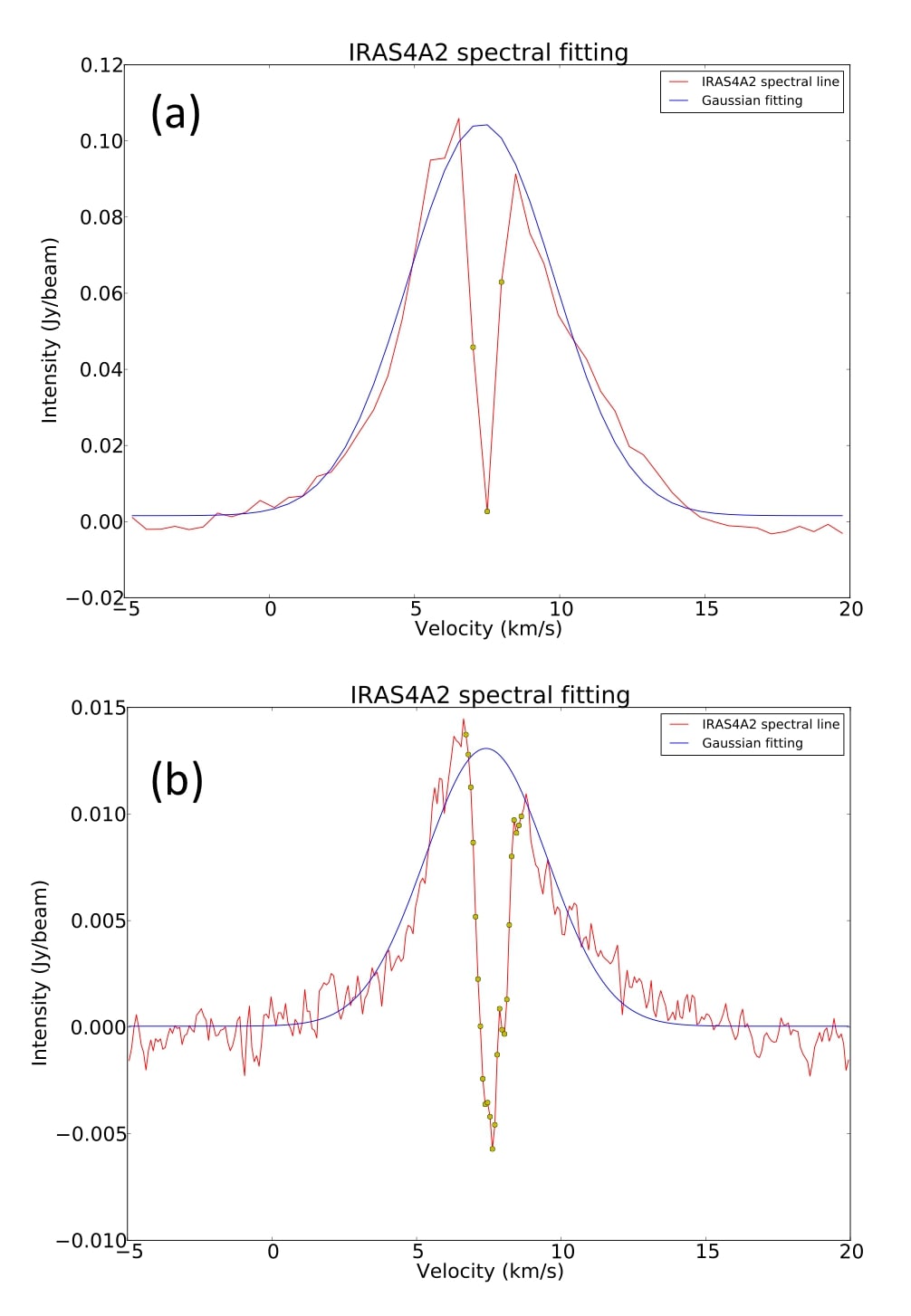}
\caption{(a) Gaussian fitting to an SO ($J_N=7_6-6_5$) line profile derived at the continuum peak of IRAS 4A2. This fitting excludes the data points below 3$\sigma$, where 1$\sigma$ corresponds to $3.6\ \mJB$. (b) Same as panel (a), but fitting to SO ($J_N=6_5-5_4$). This fitting excludes the data points below 3$\sigma$, where 1$\sigma$ corresponds to $1.3\ \mJB$. The yellow data points affected by ambient cloud absorption are ignored in the Gaussian fitting in the dip around $V=7\ \kms$.
\label{fig:slp}}
\end{figure}

As shown in the previous sections, IRAS 4A2 is surrounded by the elongated structure, the northwestern and southeastern edges of which are connected to the bases of the redshifted and blueshifted outflow lobes, respectively. In order to investigate the kinematics of this elongated component, we first determine the systemic velocity of IRAS 4A2 using the line profiles of two transitions of SO shown in Figure \ref{fig:slp}. The line profiles are derived at the continuum peak of IRAS 4A2 and fitted with a single Gaussian function using the emission above $3\sigma$, where 1$\sigma$ corresponds to $3.6\ \mJB$ and $1.3\ \mJB$ for SO ($J_N=7_6-6_5$) and SO ($J_N=6_5-5_4$), respectively. The line profile shows a dip near $V_{\rm LSR}$ = $7\ \kms$ due to the absorption by the foreground component \citep[][]{su2019}. Hence, we excluded these data in the Gaussian fitting. The peak velocity derived from the SO ($J_N=7_6-6_5$) line profile is $7.30\pm 0.05\ \kms$, while that from the SO ($J_N=6_5-5_4$) line profile is $7.40\pm 0.04\ \kms$. Averaging the two derived velocities, the systemic velocity for IRAS 4A2 is determined to be $7.35\ \kms$. This derived systemic velocity is slightly more redshifted than previous values, $6.83\ \kms$, derived from ammonia \citep[][]{Choi2010} and $7.00\ \kms$, derived from six transitions of H$_2$CO \citep[][]{su2019}.

\newpage

\begin{figure*}[ht!]
\epsscale{1.0}
\plotone{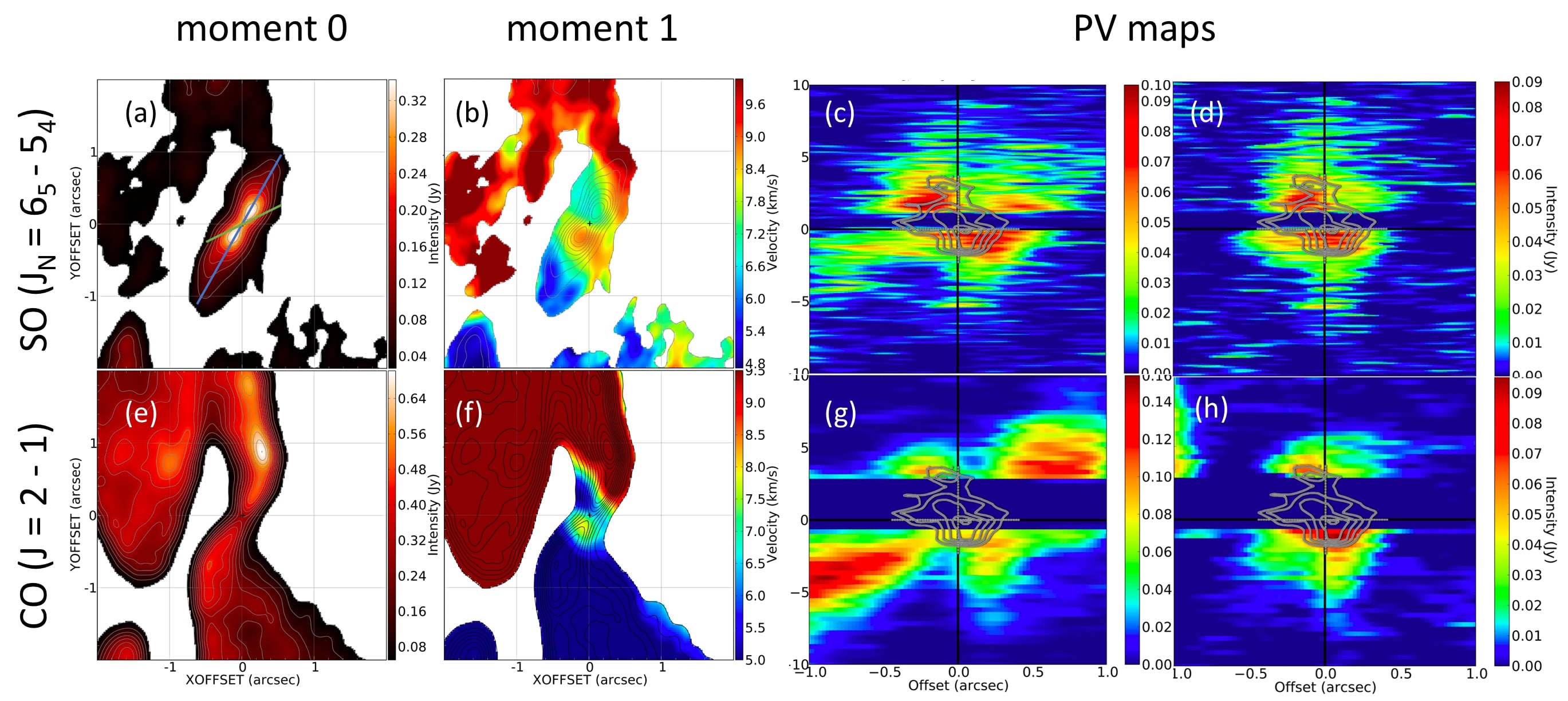}
\caption{(a) Integrated intensity map in SO ($J_N=6_5-5_4$). The integrated velocity range is from 2.4 to $13.5\ \kms$. Contour levels are $1,2,3,4,\dots \times 3\sigma$, where 1$\sigma$ corresponds to $14.2\ \mJB~\kms$. The continuum peak is located at the center. The blue and green lines are the cutting lines through the center for position-velocity diagrams of SO (length: $2 \arcsec$, P.A.: $148\arcdeg$) and ammonia (length: $1 \arcsec$, P.A.: $109\arcdeg$), respectively. (b) Same as panel (a) but mean velocity map of SO ($J_N=6_5-5_4$).  
(c) Position-velocity diagram of SO ($J_N=6_5-5_4$) (colour) derived from the major axis of SO emission (blue cut in panel (a)) and NH$_3$ ($N_J=2_2-2_2$, $F_1=3-3$ averaged with $N_J=3_3-3_3$, $F_1=4-4$) (contour) derived from the major axis of ammonia (green cut in panel (a)). The contour levels are $3,4,5,6,7 \dots \times 1\sigma$ where $1\sigma$ is 10.3 K \citep[][]{Choi2010}
(d) Same as (c), but with SO ($J_N=6_5-5_4$) PV derived from the ammonia cut. (e) Integrated intensity map of CO ($J = 2-1$). The integrated velocity range is from 2.4 to $13.5\ \kms$. Contour levels are $1,2,3,4,\dots \times 3\sigma$, where 1$\sigma$ corresponds to $14.3\ \mJB~\kms$. The continuum peak is located at the center.
(f) Same as panel (e) but mean velocity map of CO ($J = 2-1$). 
(g) Position-velocity diagram of CO ($J = 2-1$) (colour) and NH$_3$ (contour) derived along the blue cut and green cut in panel (a), respectively. 
(h) Same as (g), but with CO ($J = 2-1$) PV derived from the ammonia cut. 
\label{fig:pv}}
\end{figure*}

Figure \ref{fig:pv}a and \ref{fig:pv}b are the integrated-intensity (moment 0) and mean-velocity (moment 1) maps of SO ($J_N=6_5-5_4$) line emission, centered at the IRAS 4A2 continuum peak position. The higher resolution image of SO ($J_N=6_5-5_4$) reveals an elongated structure around IRAS 4A2 with extensions to the northwest and the southeast. This elongated structure appears to have a double peaked feature having a separation of $\sim 0\farcs 5$ ($\sim 150$ au). This double peaked feature is likely to be caused by the continuum subtraction. We confirm that the SO moment 0 map shows a single peak before continuum subtraction. This elongated structure shows a significant velocity variation along its major axis; the mean-velocity map shows a red-blue-red-blue feature from the northwest to the southeast. Within $0\farcs5$ scale, SO shows a velocity gradient from the blushifted emission in the northwest to the redshifted emission in the southeast, while the emission beyond $0\farcs5$ scale shows an opposite gradient. The higher transition SO ($J_N=7_6-6_5$) also shows the similar kinematic feature. However, the velocity gradient at the central $0\farcs5$ is not well resolved in this line due to the lower resolution of $0\farcs65$. The CO ($J = 2-1$) moment 0 and moment 1 maps in Figure \ref{fig:pv}e and \ref{fig:pv}f clearly show that the outer parts ($\pm 0.5-1\arcsec$ from IRAS 4A2 continuum peak) of the elongated structure are connected with the outflow lobes associated with IRAS 4A2, although no apparent boundary exists between the elongated structure and the outflow lobes. In addition, the velocity distribution in the outer region of the elongated structure appears to be smoothly connected to the outflow. 
This implies that the outflow dominates the velocity distribution in the outer parts. Another kinematic component could dominate the smaller scale velocity distribution because it has an opposite velocity gradient to that of the outer parts. This additional kinematic component is also reported in \citet[][]{su2019}.
The velocity gradient within the $0\farcs5$ scale is likely to be the rotation of the envelope surrounding IRAS4A2 because the observed velocity gradient from the blueshifted emission in the northwest to the redshifted emission in the southeast is consistent with that of the rotating disk traced by the ammonia emission \citep[][]{Choi2010}. It should be noted that the velocity gradient in the vicinity of IRAS4A2 observed in NH$_3$ and SO are opposite to that of the circumbinary envelope \citep[][]{Ching2016}. 

Figure \ref{fig:pv}c (SO $J_N=6_5-5_4$) and Figure \ref{fig:pv}g (CO) are the position-velocity (PV) diagrams along the line with a P.A. of $148.0\arcdeg$, which passes through the continuum peak and the double peaked feature in the SO ($J_N=6_5-5_4$) moment 0 map. These PV diagrams show two pairs of emission in the diagonal pairs of quadrants. The second and forth quadrant pair correspond to the velocity gradient of the inner parts of the elongated structure, while the first and third quadrant pair correspond to that of the outer parts. 
We interpret these two pairs of velocity components as the combination of two different motions: (1) the asymmetric double peaked pair in the second and forth quadrant represents the rotation of the elongated structure, and (2) the fainter linear pair in the first and third quadrant represents the outflow motion. 

Although the velocity gradient of the inner part of the elongated structure is interpreted as a rotation, there are several differences from the previous NH$_3$ observations \citep{Choi2010}.
First, the position angle of the major axis of the elongated structure, 149.0$^{\circ}$, is different from that of the NH$_3$ disk, 108.9$^{\circ}$.
Second, the SO PV diagram suffers from continuum subtraction near the central protostar. Third, the emission at the low velocity suffers from the foreground absorption \citep[][]{su2019} and the filtering out effect of the interferometer. Similarly, the CO PV diagram shows no emission at low velocities, which is due to filtering effect of the interferometer. 
Figures \ref{fig:pv}d and \ref{fig:pv}h compare the PV diagrams of SO and CO (color) with that of NH$_3$ (contours), which is interpreted as a Keplerian rotation of a circumstellar disk \citep{Choi2010}. The PV cuts of SO and CO are same as that of NH$_3$, $108.9\arcdeg$, which is the major axis of the ammonia moment 0 map. The PV diagrams of SO and CO generally agree with that of NH$_3$, although the rotation curve is not well traced in the SO and CO due to the missing low velocity components.

\newpage
\clearpage

\subsection{CCH ($N=3-2$) lines}
\label{sec:cch}
\begin{figure*}[ht!]
\epsscale{1.0}
\plotone{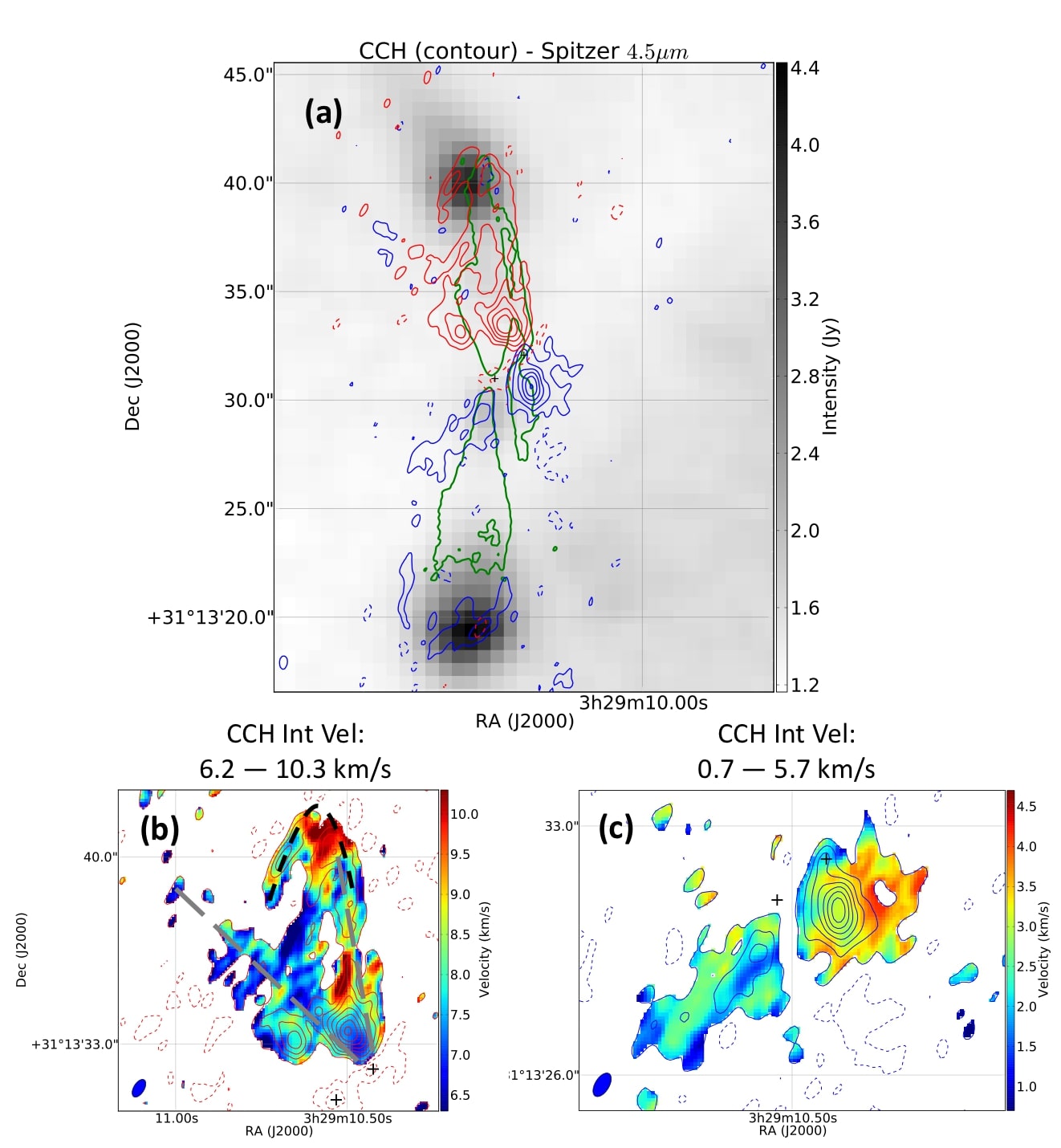}
\caption{
(a) Integrated CCH red- and blueshifted components (colored contours) overlapped with a {\it Spitzer} IRAC map at 4.5 $\micron$ (grey scale). The contour levels are $1,2,3,4,\dots \times 3\sigma$, where $1\sigma$ corresponds to 2.39 (red) and 3.34 (blue) $\mJB~\kms$. The green contour denotes the 9 $\sigma$ cutoff of CO moment 0 map in Figure \ref{fig:co}a. The continuum peak positions of IRAS 4A1 and IRAS 4A2 are denoted by black crosses. 
(b) Moment 0 (contours) and moment 1 (colour) maps for the redshifted component. Contours are the same as panel (a). Blue-filled ellipses at the bottom-left corner denote the ALMA synthesized beam: $0\farcs65 \times 0\farcs35,\ {\rm P.A.}=-28^{\circ}$. The black and grey dotted line denotes the V-shaped and U-shaped structures, respectively.
(c) Same as panel (b) but for the blueshifted component. 
\label{fig:cch}}
\end{figure*}

Figure \ref{fig:cch}a shows CCH integrated components overlapped on the {\it Spitzer} $4.5\ \micron$ map. To avoid the overlap between two hyperfine structures of CCH, we integrated the emission from 0.7 to $5.7\ \kms$ of the $F =$3--2 line for the blueshifted components, and that from 6.3 to $10.3\ \kms$ of the $F =$4--4 line is integrated for the redshifted component. The CCH also shows the redshifted component to the north and the blueshifted component to the south of the two protostars. However, the spatial distribution of CCH is significantly different from those of CO and SO; the CCH emission clearly shows two bright blobs associated with IRAS 4A2  and two faint blobs at the east of the two bright blobs, associated with IRAS 4A1. The blobs associated with IRAS 4A2 have larger sizes and higher peak intensities than those of the IRAS 4A1 blobs. The typical velocity of the IRAS 4A1 and IRAS 4A2 blobs are  $\sim \pm 3\ \kms$ and $\sim \pm 2\ \kms$, with respect to the systematic velocity of IRAS 4A1 (6.86 $\kms$, \citet[][]{su2019}) and IRAS 4A2 (7.35 $\kms$), respectively. 
The two IRAS 4A2 blobs exhibit clear bipolarity with an extension of P.A.$=23\arcdeg$. This position angle is different from the extension of the CO and SO outflows, $\sim$0$\arcdeg$, in the central 3$\arcsec$ region of IRAS 4A2. The extension of the CCH outflow is more perpendicular to the major axis of the SO elongated structure around IRAS 4A2 than CO and SO outflow (i.e. compare CCH and CO moment 0 maps in Figure \ref{fig:cch}a). To the north and south of these blobs, CCH emission shows U-shaped structures with their tips pointing away from IRAS 4A1 and IRAS 4A2 continuum peaks. These U-shaped component coincide with the bright components in {\it Spitzer} $4.5\ \micron$ emission. The $4.5\ \micron$ emission represents H$_2$ regions, which usually traces shocked regions \citep[e.g.][]{Santangelo2015}. This indicates that the U-shaped feature in CCH might also trace a shocked region. Since the bright components in {\it Spitzer} $4.5\ \micron$ emission are assumed to be associated with IRAS 4A1, the U-shaped feature in CCH emission is also associated with IRAS 4A1. Figure \ref{fig:cch}b and \ref{fig:cch}c show that each blob except the south-western blob is connected with a fainter extension toward the north and the south. The fainter extension consists of a western edge and an eastern edge, forming a V-shaped structure with its tip pointing toward IRAS 4A2 continuum peak. The western edge of the V-shaped structure is more redshifted than the eastern edge of that. The southern blob of IRAS 4A2 also shows the same velocity gradient. This velocity gradient is identical to the west-to-east velocity gradient in the circumbinary envelope and the velocity gradient in SO very low velocity component. 
\newpage

\begin{figure}[ht!]
\epsscale{1.0}
\plotone{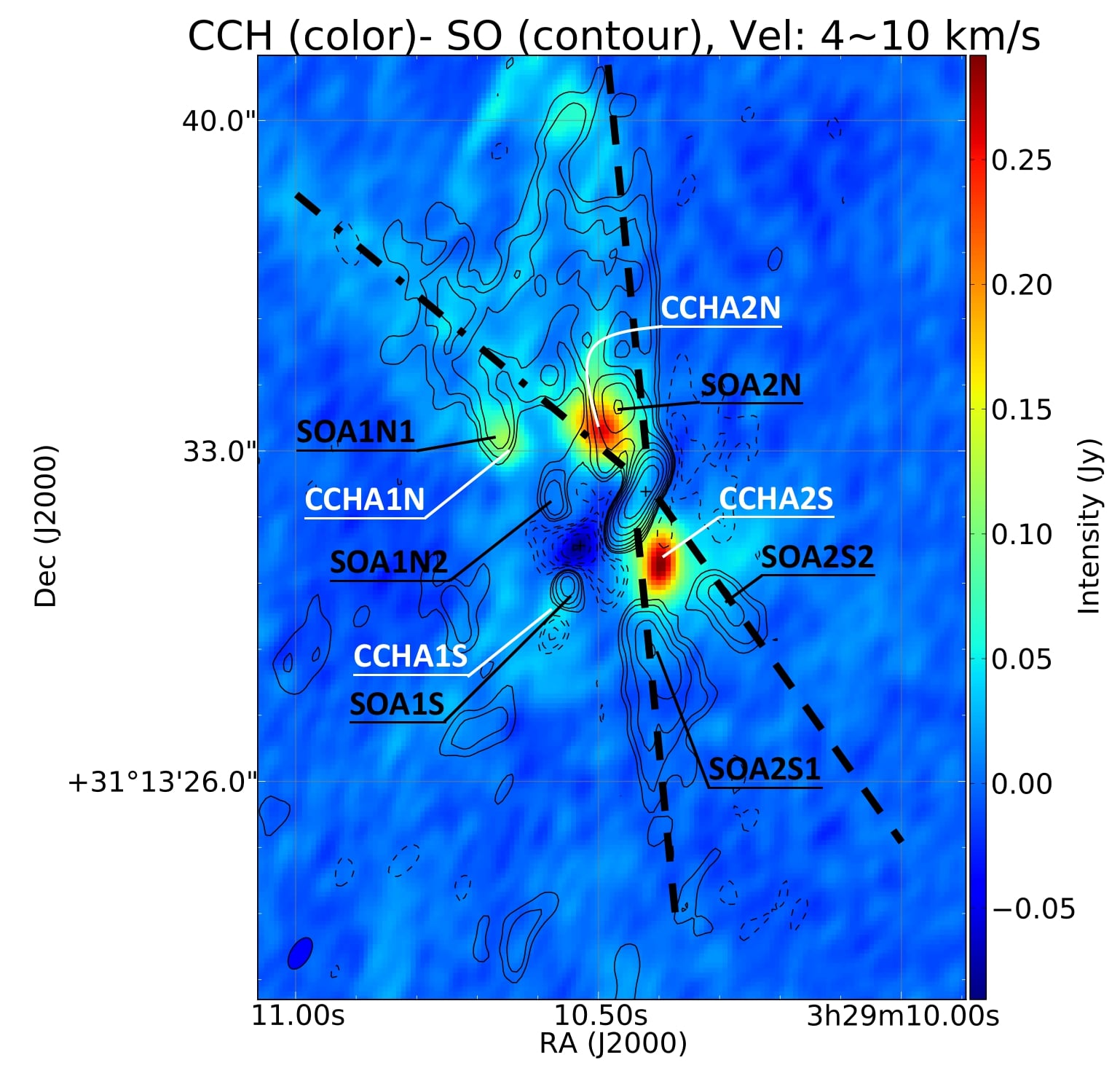}
\caption{CCH integrated intensity (contours) overlapped with SO very low velocity component (colour). To avoid the overlap between two hyperfine structures in CCH, the emission from $5.7\ \kms$ to $6.3\ \kms$ is removed. The contour levels are $1,2,3,4,\dots \times 3\sigma$, where $1\sigma$ corresponds to 5.0 $\mJB~\kms$. The continuum peak positions of IRAS 4A1 and IRAS 4A2 are denoted by black crosses. The white and black texts denote the name of each blob shown in CCH and SO emission, respectively. Blue-filled ellipse at the bottom-left corner denote the ALMA synthesized beam: $0\farcs65 \times 0\farcs35,\ {\rm P.A.}=-28^{\circ}$.
\label{fig:cchso}}
\end{figure}

While both overall CO and SO emission (Figure \ref{fig:co}a and Figure \ref{fig:so}a) show an asymmetric S-shape in the IRAS 4A2 outflow, the CCH outflow shows a symmetric feature which extends along northeast to southwest direction. However, in IRAS 4A2, the SO emission at the very low velocity shows similar feature to CCH outflow. Figure \ref{fig:cchso} shows the very low velocity SO moment 0 map (contour) overlaid by the CCH (color) moment 0 map with the same integrated velocity. The CCH and the very low velocity SO emission extended in the same northeast-to-southwest direction, while both of the line emission show blob components near the IRAS 4A2 continuum peak. The northern CCH blob of IRAS 4A2 (CCHA2N in Figure \ref{fig:cchso}) overlaps with the northern SO blob (SOA2N) at the very low velocity, with the intensity peak of CCH blob slightly shifted to the west of the SO blob. Meanwhile, the southern CCH blob of IRAS 4A2 (CCHA2S) is at the north of the two southern SO blobs (SOA2S1 and SOA2S2). The V-shaped component in northern CCH outflow traces the two edges of the triangular component in northern SO outflow. The northeastern edge of the V-shape is unclear due to the overlap of IRAS 4A1 and IRAS 4A2 lobes. Accordingly, the CCH and SO low velocity emission enhances at the same region in northern IRAS 4A2 outflow lobe. Assuming that the very low velocity SO outflow share the same P.A. with CCH outflow in IRAS 4A2, the extension of the very low velocity SO outflow is elongated along the same direction as the CO and SiO  \cite[][]{Ching2016, Santangelo2015} while misaligned with the SO high velocity outflow.
Additionally, the two fainter blobs in CCH (CCHA1N and CCHA1S) associated with IRAS 4A1 are also traced by the low velocity SO emission (SOA1N1 and SOA1S). There is one additional faint blob (SOA1N2) in the SO moment 0 map which is not seen in the CCH moment 0 map near the continuum peak of IRAS 4A1. 

\newpage

\subsection{Main Results About the IRAS 4A2 Outflow-Envelope System}
Here we summarize the main results in IRAS 4A2 with the high resolution observation of CO and SO line emission. 
\begin{enumerate}
\item IRAS 4A2 drives its own bipolar outflow toward the north (redshifted) and the south (blueshifted), respectively.

\item The IRAS 4A2 outflow at different velocity ranges show different morphologies, i.e. the very low velocity outflow shows a symmetric cone shape while the low velocity outflow shows an asymmetric S-shape.

\item The low velocity asymmetric S-shape outflow consists of two curved outflow lobes extended to the P.A. of 20$\arcdeg$ and an elongated structure along the P.A. of 148$\arcdeg$ at the center of the S-shape. 

\item Extremely high velocity components at the base of the IRAS 4A2 outflow is observed in CO emission.

\item The velocity gradient within $\pm 0\farcs5$ of the elongated structure, which is considered to be the envelope of IRAS 4A2, is opposite to that of the circum-binary envelope. 

\item The PV diagrams of CO and SO line emission show two pairs of velocity components, which corresponds to the rotation of the elongated structure and outflow motion. 

\end{enumerate}
In order to explain the observational features, we compare our results with the models in Section \ref{sec:mech}.

\section{Possible Mechanisms for S-shaped Outflow Associated with IRAS 4A2}
\label{sec:mech}
In this section, we will focus on the S-shape outflow associated with IRAS 4A2 and attribute it to three possible mechanisms, rotation of the outflow, the outflow precession caused by a close binary, and the misaligned B-field and cloud initial rotation axis.

\subsection{Rotation of the Outflow}
\label{sec:outrot}
A rotating outflow could be the origin of the S-shaped outflow. The rotation of outflow lobes shifts the radial velocity of one side of the outflow cavity close to the systemic velocity, which can be resolved out by the interferometer, leaving the opposite side of the outflow cavity. Since this resolving out would occur at different sides of the blueshifted and the redshifted outflow cavities, the observed outflow shows an asymmetric S-shape morphology. In such a case, the outflow is expected to be rotating in the same direction as the protostellar disk, and has a specific angular momentum comparable with that of the disk.
However, this scenario is unlikely, because the required outflow rotation, which has an approaching side in the east and a receding side in the west, is opposite to the rotation of the disk traced by ammonia \citep[][]{Choi2010}.

\subsection{Outflow Precsession Caused by Close Binary}
\label{sec:prec}
Close binaries could reproduce curved outflows \citep[e.g.][]{bate2000,hirano2010,kwon2015}, which are similar to the S-shaped outflow associated with IRAS 4A2. So far, none of the existing continuum images shows a signature of binary in IRAS 4A2. The upper limit of the projected binary separation is estimated from the beamsize of the highest resolution image of \citet[][]{tobin2016}, i.e. 0\farcs{075} $\times$ 0\farcs{05} (22.4 au $\times$ 14.7 au). Then, we tried to model the outflows launched from precessing binaries separated by different distances (from 5.9 au ($0\farcs02$) to 22.4 au ($0\farcs075$)). We found that while a model with a separation larger than 12 au ($0\farcs04$) will fit the observed S-shape outflow, that with a separation lower than 12 au ($0\farcs04$) will start to deviate from the observed S-shape. Here, we present a succeeded model of 14.7 au separation and a failed model of 8.8 au separation. In the models, the precession period is 20 times longer than the orbital period of the binary \citep[][]{bate2000}. Assuming a Keplerian motion of the binary, the precession period is derived by the equation $T_{p} {\sim} 20\times T_{o} = 20\times\sqrt{\frac{4\pi^2D^3}{G\times M_{A2}}}$, where $T_p$ is the precessing period, $T_o$ is the orbital period, $D$ is the distance between the close binary, $G$ is the gravitational constant, and $M_{A2}$ is a total mass of the binary system, which is estimated to be $0.08\ M_\odot$ in \citet[][]{Choi2010}. The derived precessing period is 4000 years. We build the precession models in a 3D spherical coordinate system, and the precession axes of the models coincides with the zenith direction in the coordinate system. A positive $\theta$ value corresponds to the polar angle with respect to the zenith direction, and a positive $\phi$ value corresponds to the counter-clockwise azimuthal angle looking from positive zenith direction.

To model the outflow trajectory, two outflow velocity patterns are considered in the precession models. One is a constant outflow velocity at every positions of the outflow \citep[][]{kwon2015}. The free parameters we fit in this model are the viewing angle, half the opening angle of precession ($\alpha$), and the outflow axial velocity ($v$).
\begin{equation} \label{eq:3}
\begin{Bmatrix} 
r=vt\\
\theta=\alpha\\
\phi = \frac{Nt}{T_{p}}\times 2\pi
\end{Bmatrix}
\end{equation}
Where $N$ is $+$1 for counter-clockwise rotation and is $-$1 for clockwise rotation of the precessing binary, and $t$ is the traveling time of the outflow. We adopted the timescale of the model to be 6000 year, which is the outflow dynamical timescale estimated in \citet[][]{yildiz2012}.

Another outflow velocity pattern is an outflow with its velocity increasing as a function of the distance \citep[e.g. the ``wind driven model"][]{lee2000}. The accelerating outflow is described by the equation $\frac{dv_z}{dz}=v_0$ in this model, and $v_0$ is estimated by the boundary conditions below: (1) $0\ \kms$ at outflow launching point ($r=0\arcsec$), (2) the highest radial velocity achieved by the IRAS 4A2 outflow, $V_{LOS}\sim5.0\ \kms$, at $\sim7\arcsec$ away from the outflow launching point. The highest 3D velocity of the outflow need to be corrected by the outflow inclination. 
Since \citet[][]{yildiz2012} and \citet[][]{choi2006} suggested  different inclination angles of $45\arcdeg - 60\arcdeg$ and $79.3\arcdeg$ with respect to the line of site, respectively, we constructed the models for both cases with 52.5\arcdeg and 79.3\arcdeg. Thus, the $v_0$ solved by $V_{LOS}/cos(52.5\arcdeg)=8.2\ \kms$ and $V_{LOS}/cos(79.3\arcdeg)=27\ \kms$ resulted in $v_0=4.0\times10^{-3}\ \kms \ au^{-1}$ and $v_0=13.1\times10^{-3}\ \kms\ au^{-1}$, respectively. Solving $\frac{dv_z}{dz}=v_0$, we obtained the velocity of the outflow $v_z$ as a function of the outflow traveling time $t$, $v_z=Ce^{v_0t}$ and the outflow traveling distance $r$ as a function of time, $r=\frac{C}{v_0}e^{v_0t}-\frac{C}{v_0}$ (we assume $r=0$ when $t=0$), where $C$ is a constant. For accelerating outflow model, the free parameters to be fitted are the viewing angle, the half opening angle of the precession, and the constant $C$.
\begin{equation} \label{eq:4}
\begin{Bmatrix} 
r=\frac{C}{v_0}e^{v_0t}-\frac{C}{v_0}\\
\theta=\alpha\\
\phi = \frac{Nt}{T_{p}}\times 2\pi
\end{Bmatrix}
\end{equation}

The 3D model is then projected on the plane of the sky. The free parameters are fitted so that the model trajectory to align with the intensity peak positions of the moment 0 map of CO emission, while the model line of site velocity to align with that of the CO emission at the peak positions. The intensity peak positions are estimated from the Gaussian fitting of each row of pixels in the moment 0 map. We choose the peak positions of the fitted Gaussian to be the outflow intensity peaks, and the $\sigma$ of the Gaussian to be the error of the peak positions. The emission associated with IRAS 4A1 outflow is masked out manually during the fitting. The value and error of the line of site velocity is the mean velocity (moment 1) and the velocity dispersion (moment 2) in CO emission, respectively.

\begin{figure}[ht!]
\epsscale{1.0}
\plotone{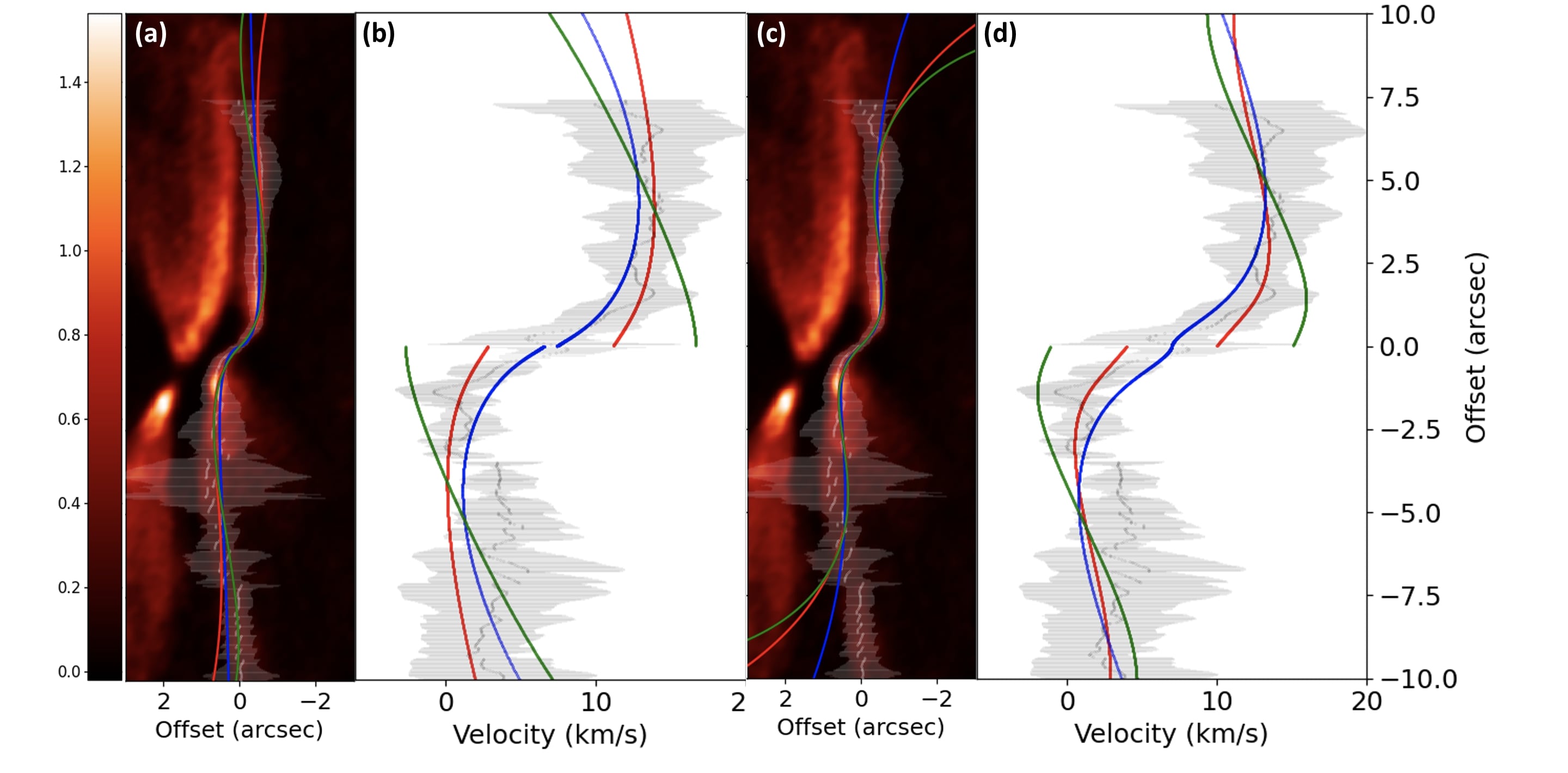}
\caption{(a), (c): CO moment 0 maps (integrated intensity, color) overlaid with the outflow precession model (red, blue and green lines). Panel (a) and (c) show the best fit outflow curve with a binary separation of 14.7, and 8.8 au, respectively. The red and blue lines are the best fit accelerating outflow curve with outflow inclination of 52.5$\arcdeg$ and 79.3$\arcdeg$ with respect to the line of site, respectively. The green lines are the best fit constant velocity outflow curve. The dotted lines and the white regions are the intensity peaks of CO emission from IRAS4A2’s outflow and it's error. (b), (d): Mean velocity (moment 1, grey dots) and velocity dispersion (moment 2, grey region) overlapped with the line of site velocity of the precession model (red, blue and green lines). Panel (b) and (d) show the best fit outflow curve with a binary separation of 14.7, and 8.8 au, respectively. The red, blue and green lines are the same as those in panel (a) and panel (c), but showing the line of site velocity of the precession model.
\label{fig:precession_1}}
\end{figure}

Figure \ref{fig:precession_1} shows the results of the fitting of the precession models for the binary separations of 14.7 au and 8.8 au. The model with a separation of 14.7 au can explain the overall morphology of the S-shaped outflow for both constant velocity and accelerating outflows. On the other hand, in the models with a narrower separation of 8.8 au, the the model trajectories starts to deviate with the observed S-shape. The observed velocity pattern is better explained with the accelerating outflow models for both 14.7 au and 8.8 au cases. It should be noted that no precession model could reproduce the red-blue red-blue feature in the vicinity ($\sim\pm2\arcsec$) of IRAS 4A2. The precession model can overall reproduce the S-shape outflow morphology if the binary separation is comparable to the beamsize of the highest resolution image, i.e. 14.7 au, and the outflow velocity increases as it travels. However, the precession models cannot explain the small scale kinematics. In order to search for the signature of binary in IRAS 4A2, higher resolution observations are necessary. Additionally, perturbations due to an orbital binary motion is not considered in our precession model because these perturbations should occur at a scale smaller than 100 au \citep[$\sim 0\farcs3$,][]{raga2009}, which is smaller than the beam size of our observation.

\newpage

\subsection{Magnetic Field Misalignment Model} 
\label{sec:analysis}
Next, we examine the MHD simulations by  \citet[][]{Hirano2019, machida2020} with misaligned B-field and cloud initial rotation axis, whose resultant density distribution shows potential to reproduce the S-shape feature. 

\subsubsection{The Numerical Simulation} 
\label{sec:bmam}
Previous observations of IRAS 4A2 show that the position angles of the large scale magnetic field \citep[$\sim45\arcdeg$,][]{Matthews2009, attard2009, hull2014} and moderate scale magnetic field \citep[$61\arcdeg$,][]{Ching2016, hull2014} are not aligned with the rotation axis of the Keplerian disk \citep[$19\arcdeg$, ][]{Choi2010}. We, therefore, adopted the resistive MHD simulation of \citet[][]{Hirano2019} and \citet[][]{machida2020}, which simulated the evolution of the protostellar core with an angular momentum vector tilted from the magnetic field. This misalignment results in an S-shaped configuration of density distribution, because the collapse on large scale is dominated by the magnetic field, while that on small scale is dominated by the rotation. The initial magnetic field of this simulation is set in the direction of $\it{z}$ axis, and the initial rotation axis is inclined by 30$\arcdeg$, 45$\arcdeg$, 60$\arcdeg$, and 80$\arcdeg$ with respect to $\it{z}$ axis on the $\it{x}-\it{z}$ plane. The coordinate system of this simulation is a right-handed Cartesian system. We adopted the same initial condition as the simulation done in the paper. The initial temperature and number density are 10 K and 1.5$\times$10$^4$ cm${^-}{^3}$, respectively. The simulation time of the model we used is 500 years after the time when the collapsing center reaches $\rm{n_H}$ = $10 ^{18}\ \rm{cm}^{-3}$. Other details and simulation results are shown in \citet[][]{Hirano2019}.

In order to compare the results of the MHD simulations with the observed SO image, we utilize Radmc3d radiative transfer code \cite[][]{radmc3d2012}.
We set the SO abundance relative to H$_2$ abundance to $10^{-9}$ \citep[][]{Lee2010}. A continuum image with bandwidth of 4 GHz centered at the rest frequency of SO ($J_N=6_5-5_4$) (219.949442 GHz) is also created to conduct continuum subtraction of the line image. 
Then, using the task vis\_sample (For further references see \href{https://github.com/AstroChem/vis_sample.git}{vis sample}\footnote{https://github.com/AstroChem/vis\_sample.git}), we simulate the observation of the unsubtracted line image with the same UV coverage of our observation. 

We fine tune the initial rotation axis and viewing angle to identify the synthetic result which looks most similar to the observational image. The initial rotation axis is inclined by 60$\arcdeg$ with respect to $\it{z}$ axis, and the viewing angle is 121$\arcdeg$ inclined from $\it{z}$-axis, clockwise 68$\arcdeg$ from $\it{x}$-axis, and a position angle of 35$\arcdeg$.
Last, the continuum is subtracted from the visibility created by the task vis\_sample through the CASA task {\it uvcontsub}, and CLEANed by the task {\it tclean} with natural weighting. 

\subsubsection{Simulation Results} 
\label{sec:simrslt}
The results of the MHD simulation are presented in \citet[][]{Hirano2019} and \citet[][]{machida2020}. The simulated density profile shows that the normal of the inner disk and the outflow launched from the inner disk are roughly parallel to the initial rotation axis, while the outer flattened envelope is warped. The outflow launched from the inner disk overlaps with part of the large envelope and one side of the each outflow lobe is enhanced as well.

\begin{figure*}[ht!]
\epsscale{1.2}
\plotone{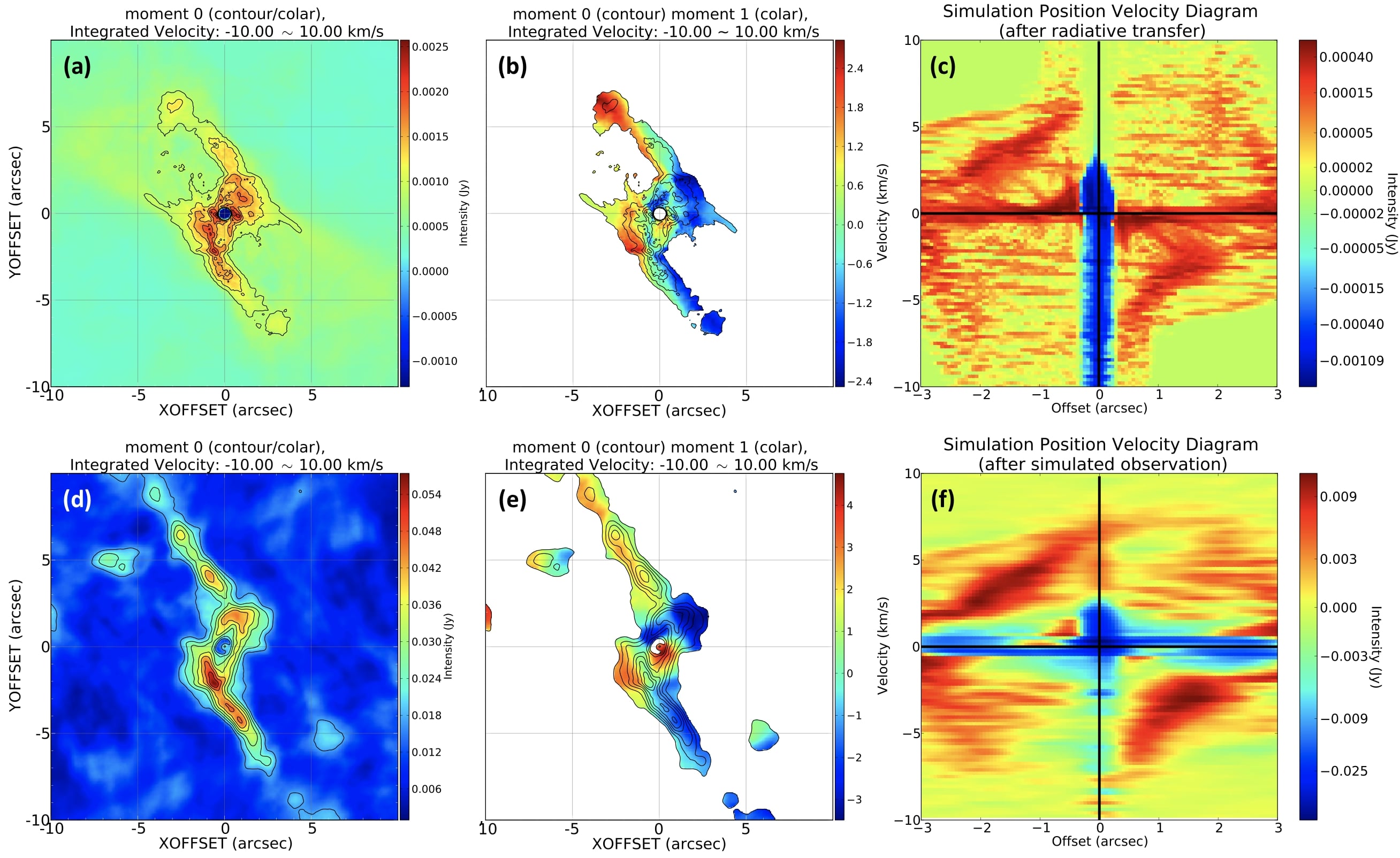}
\caption{Results after solving the radiative transfer (upper row) compared with results after performing the synthetic observation (lower row), where the left, middle, and the right columns are moment 0, moment 1, and position-velocity diagrams, repectively. The integrated velocity ranges are showed above all panels. The contour levels are  $1,2,3,4,\dots \times 3\sigma$, where 1$\sigma$ corresponds to  $3.9\times 10^{-4}\ $Jy $\kms$ in panels (a) and (b), $9.4\times 10^{-3}\ \JB~\kms$ in panels (d) and (e). The position-velocity diagrams are derived along the cut centered at continuum peak with the position angle of $110\arcdeg$ and the length of $3\farcs0$, whereas the colour bars are adjusted to highlight the outflow components.
\label{fig:smu}}
\end{figure*}

The panels in Figure \ref{fig:smu} present the misalignment model after solving the radiative transfer (the first row), and that after the synthetic observation (the second row). The first, second, and third columns are moment 0 maps, moment 1 maps, and PV diagrams, respectively. All PV diagrams are derived from the cutting line through the center of moment 0 and 1 maps with the P.A. of 110$\arcdeg$, which is the major axis of the remaining emission from the flattened envelope in the moment 0 map after synthetic observation (see below).

The simulation after solving the radiative transfer shows an asymmetric feature in moment 0 map (Figure \ref{fig:smu}a). This feature consists of a $\sim5 \arcsec$ scale elliptical component and a $\sim15 \arcsec$ scale S-shaped component, representing the flattened envelope and the outflow wall, respectively. Moment 1 map (Figure \ref{fig:smu}b) shows that the northern outflow is redshifted and the southern outflow is blueshifted. The flattened envelope has a velocity gradient from west (blueshifted) to east (redshifted). The PV diagram shows emission in all four quadrants. 
The emission in the second and forth quadrants corresponds to the rotation of the flattened envelope, and that in the first and third quadrants corresponds to the outflow. The center of the PV diagram shows absorption due to effects of continuum subtraction.

After the synthetic observations, the moment 0 map (Figure \ref{fig:smu}) shows a $\sim15 \arcsec$ scale asymmetric S-shape feature. Some of the low velocity emission in the extended flattened envelope is filtered out, and only the emission from the compact region is remaining. The compact component shows double peaks at northwest and southeast with a fainter dip at its center. There are also two spikes attached to the northeast and southwest of the compact component. The moment 1 map (Figure \ref{fig:smu}e) shows a red-blue red-blue feature along the S-shaped feature from the northwest to southeast across the central source. The PV diagram (Figure \ref{fig:smu}f) also shows two pairs of emission in the diagonal pairs of quadrants. The center of the PV diagram shows absorption, which also appears in the PV map before synthetic observations.

The simulation successfully reproduces three main features in the observation.
First, the S-shaped feature is reproduced by means of the intensity enhancement in the western side of the northern lobe and the eastern side of the southern lobe. Second, the mean velocity in the simulated moment 1 map shows the same pattern as that in the observed moment 1 map. Third, the simulated PV diagram also shows rotation and outflow components. 

However, the simulation result also differs from the observation; the simulated moment 0 map shows spike like components which correspond to the bases of the eastern and western walls of the redshifted and blueshifted cavities, respectively. In addition, the elongation of the flattened envelope is not aligned with the outflow. The differences above could be yielded by the following assumptions which are not very realistic. Our simulation assumes the SO molecule to be uniformly distributed, while in reality, the SO abundance could be enhanced under certain conditions such as outflow shells \citep[][]{aso2018}, jets \citep[][]{lee2010}, and the outer edge of disks \citep{ohashi2014}. Because the temperature dependence of the SO abundance in our simulation is too simplified, some additional features could appear in the simulation. In addition, our simulation does not include the binary companion and the foreground ambient cloud, since it is too complicated to take these components into account. 

\newpage

\begin{figure}[ht!]
\epsscale{1.0}
\plotone{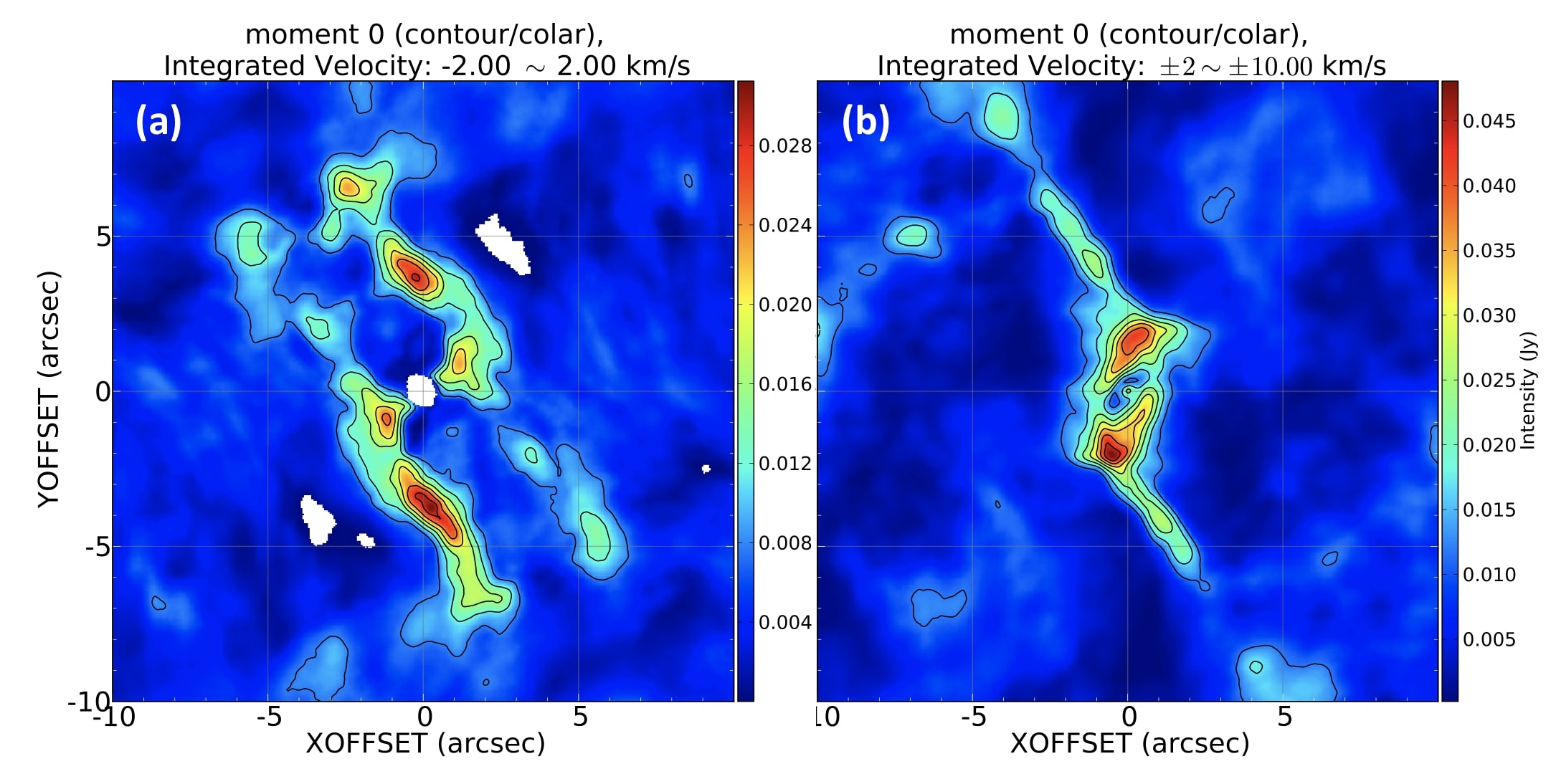}
\caption{Moment 0 maps with integrated velocity ranges of (a) $0.0-\pm 2.0\ \kms$ and (b) $\pm 2.0-10.0\ \kms$. The contour levels are  $1,2,3,4,\dots \times 3\sigma$, where 1$\sigma$ corresponds to (a) $4.5\times 10^{-3}$ and (b) $5.4\times 10^{-3}\ \JB~\kms$. 
\label{fig:smu_comp}}
\end{figure}

In order to explore the outflow morphology in different velocity ranges, such as the symmetric morphology in the very low velocity range and the asymmetric morphology at higher velocity range in the SO emission, we compare the simulated moment 0 maps with different integrated velocities. Figure \ref{fig:smu_comp}a and Figure \ref{fig:smu_comp}b show the moment 0 maps at a lower and a higher velocity range, where the integrated velocity ranges are $V = 0.0 - \pm 2.0$ and $\pm 2.0 \sim \pm 10.0 \kms$, respectively. The lower velocity component shows an X-shape at a $15\arcsec$ scale, which is more extended and less asymmetric than the higher velocity S-shaped component. This difference between the high and low velocity components is similar to that observed in the SO line, i.e. the outflow is symmetric rather than S-shape in the very low velocity (Figure \ref{fig:so}b), while the S-shape is prominent in the low velocity (Figure \ref{fig:so}c). One explanation of this difference is that the high velocity outflow is launched by outer edge of Keplerian disk (inner part of the flattened envelope) and the low velocity outflow is launched by the outer part of the flattened envelope. Since the flattened envelope is warped, the Keplerian disk and the outer part of the flattened envelope are misaligned. As a result, the high velocity outflow is misaligned with the low velocity outflow \citep[][]{Hirano2019}.

\newpage

\section{DISCUSSION} \label{sec:discussion}
\subsection{A Schematic View of the Magnetic Field Misalignment Model} 
\label{sec:bend}
\begin{figure*}[ht!]
\epsscale{1.2}
\plotone{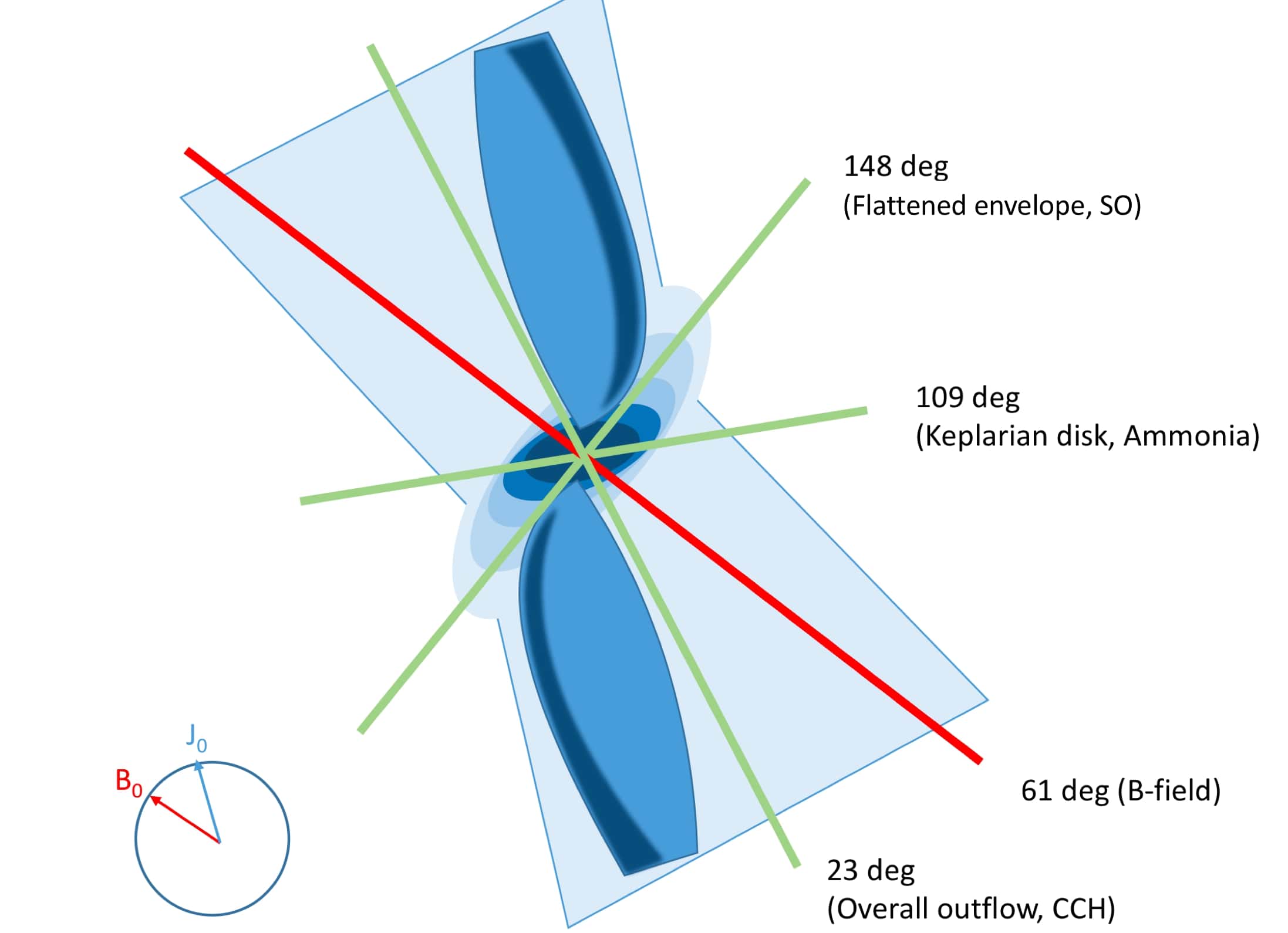}
\caption{
A schematic view of the B-field misalignment model. The red line represents the initial B-field direction \cite[][]{Ching2016}. The dark blue components show the central rotation-dominated area and a collimated high speed outflow ejected from the center. The faint blue components show the outer B-field-dominated area of the flattened envelope and an extened low speed outflow ejected from the outer part of the flattened envelope. The black shaded regions denote the interaction area of low and high speed outflows. The interaction between the outflows with different ejection angles results in an edge enhanced outflow wall (black shaded), and apparently looks like an outflow misaligned with the normal of the flattened envelope. The red and blue arrows in the bottom left circle denote the initial magnetic field and initial angular momentum vector of the simulation. The initial angular momentum vector is assume to be perpendicular to the ammonia disk elongation. 
\label{fig:sch}}
\end{figure*}

The misalignment model can overall reproduce the outflow misaligned with the flattened envelope. Figure \ref{fig:sch} is a schematic view of this model. It tells us that while the inner velocity gradient mostly depends on the initial rotation, the outer part is shaped more by the magnetic field (or Lorentz force). This results in two features. First, it is more efficient for materials to fall along the magnetic field rather than across the magnetic field, so the flattened envelope is warped due to the different orientation between rotation dominated and B-field dominated regions. The outer parts of the envelope mainly flatten in the direction perpendicular to the magnetic field, while the central part of the envelope, which also corresponds to outer edge of Keplerian disk, contracts along the initial rotation vector. Consequently, the velocity gradient in a large scale is misaligned with that in a smaller scale due to the misaligned orientation of the large and the small scale. Secondly, the outflow materials launched from different parts of the disk show different orientations. The high velocity outflow launched from the outer edge of the Keplerian disk is ejected along disk rotation axis, while the slow extended outflow launched from the outer part of the flattened envelope is ejected along the direction between B-field and initial rotation, aligned with the B-field flattened envelope. As a result, the whole flattened envelope shows a warped structure and the two different outflow components show different extension. The strong high velocity outflow enhanced the density of one edge in the extended outflow, and thus looks like an S-shaped outflow misaligned with flattened envelope rotation. In this perspective, we expect to see two features in this model. One is that the velocity gradient of the Keplerian disk is misaligned with the larger scale velocity gradient. This is presented in ammonia moment 0 maps, showing that the major axis of a small scale Keplerian disk differs from that of large scale SO flattened envelope. The other is that the outflow launched from the outer edge of Keplerian disk is faster and more aligned with the rotation axis of the Keplerian disk than that from the flattened envelope. This is shown in both observation and simulation when we compare the moment 0 maps with the high and the low integrated velocity (Figure \ref{fig:so}). According to \citet[][]{Hirano2019}, the high speed outflow in the simulation is collimated and shows knot like components. While the collimated outflow is not obvious in our SO line observation, the CO observation shows extremely high velocity components at the base of the outflow, which could correspond to the high speed outflow in the simulation. 

In addition, MHD simulations \citep[e.g.][]{Hirano2019, ciardi2010, matsumoto2004} suggest that misaligned configuration of B-field and cloud angular momentum vector could be the origin of precession, which changes the direction of jet/outflow as a function of time.
This could further explain the larger scale S-shaped pattern from P.A. $\sim20\arcdeg$ at $r<20\arcsec$ to P.A.$\sim45\arcdeg$ at $r>20\arcsec$ \citep[e.g.][]{yildiz2012,choi2005}. 

While we explain the S-shaped outflow with misaligned B-field and cloud angular momentum vector, the origin of the misalignment is still unclear. A possible mechanism of producing such misalignment is turbulent fragmentation \citep[][]{goodwin2004I, goodwin2004II, goodwin2006}. The angular momentum of the local parts in IRAS 4A cloud is independent from the large scale B-field in this mechanism. As a result, when the large cloud collapses and fragments into IRAS 4A1 and IRAS 4A2, the angular momentum vector of IRAS 4A2 could be misaligned from the large scale B-field, which then induced subsequent features such as S-shaped outflow. 

\newpage

\subsection{Possible Mechanisms of the Opposite Velocity Gradient} 
\label{sec:pmovg}
The velocity gradient of the flatten envelope traced by SO emission is consistent with that traced by the ammonia \citep[][]{Choi2010}, while opposite from that of the circumbinary envelope traced by C$^{17}$O \citep[][]{Ching2016}. Here, we discuss two possible mechanisms for such opposite velocity gradient.

One possible mechanism to change the velocity gradients is the turbulence in magnetized molecular cloud cores. Recently, several ideal MHD simulations of magnetized-turbulent molecular cloud cores were conducted \cite[e.g.][]{myers2013, joos2013, seifried2013, matsumoto2017}, showing that the large scale envelope rotation and small scale disk rotation could be misaligned. If the turbulence is strong enough, it could cause the small scale disk angular-momentum vector to be largely misaligned with large scale rotation and thus shows a nearly opposite velocity gradient between large and small scales \cite[][]{takakuwa2018}. In addition, the turbulence also contributes to the misalignment of disk rotation axis and large scale magnetic field, which latterly tilt the outflow direction from the disk rotation axis \cite[][]{matsumoto2017}. Although these features are consistent with the observation toward IRAS 4A2, these misalignment only take place in the scale smaller than $\sim 100$ au, so it is less possible to cause the opposite velocity gradient between a $\sim 300$ au flattened envelope and an even larger circumbinary envelope.

On the other hand, in non-ideal MHD cases, Hall term in the induction equation induces a toroidal magnetic field on the cloud core. When the initial magnetic field is anti parallel to the cloud core angular momentum vector, the Hall induced magnetic field would run opposite to the gas rotation. Mean while, Hall effect also exerts a magnetic torque on the mid-plane of the flattenned envelope. If this torque is large enough and have opposite direction of initial rotation, the gas rotation vector could flip between the large scale core and the small scale flattened envelope. On the other hand, it is also possible for counter rotation between large and small scale components when the magnetic field is parallel to the disk rotation vector. In this case, the Hall induced magnetic torque produces an excessive angular momentum on the mid-plane of the disk. However, due to the conservation of angular momentum, a negative angular momentum is transferred to the upper region of the disk and form a counter rotating envelope. Recently, it is also found that the Hall induced counter rotation could appear in cases other than anti-parallel or parallel configuration (i.e. magnetic field and disk rotation vector misaligned by 135$\arcdeg$), so the counter rotation could happen in the misaligned source IRAS 4A2 \cite[][]{tsukamoto2017}. Such an opposite velocity gradient, possibly caused by Hall effect, is reported in IRAS 04169+2702 and L1527 \cite[][]{takakuwa2018, harsono2014}. In the case of IRAS 04169+2702, the opposite velocity gradient is observed with $^{13}$CO and C$^{18}$O line emission. By fitting a counter-rotating model and a forward rotation model to $^{13}$CO and C$^{18}$O position-velocity diagrams, it is found that counter-rotating model better explains the opposite velocity gradient. 

\subsection{Increase Abundance in CCH} 
\label{sec:iac}
The spatial distribution of the CCH is significantly different from those of the CO and SO; although the CCH also reveals a bipolar structure, it consists of two blobs at north and south of IRAS 4A2. 
Normally, CCH appears on outflow cavity walls because its raw material, some ions like C$_2$H$_2^+$ or C$_2$H$_3^+$, are formed there. These ions are the leftovers of FUV light dissociation of some carbon bearing neutral molecules. When FUV light from protostars shed on the inner wall of cavities, neutral molecules will break down into ions thus start to form CCH on cavity walls. A typical source demonstrating CCH traced cavity wall is NGC1333 IRAS4C. Since CCH can be formed in PDRs irradiated by FUV radiation \cite[e.g.][]{pety2005, zhang2018}, it is interpreted that the observed CCH is tracing the outflow cavities irradiated by the FUV radiation from the central star. However, in contrast to IRAS 4A, the molecular outflow from IRAS4C is barely seen in the CO emission; only weak and compact blueshifted emission was detected to the east of the central source \cite[][]{stephens2018}, despite the clear cavity structure in the CCH and mid-IR images.
However, in addition to cavity, there are also CCH clumps at the bottom of cavities associated with IRASA2. Although the morphology of the CCH in IRAS 4A2 is different from that of IRAS4C, the same mechanism of CCH formation in PDRs could also be applied to this case. Since the velocity of CCH in IRAS 4A2 is much lower than those of CO and SO, the observed CCH clumps could be the bases of the cavities. The different morphologies, i.e. clumps in IRAS 4A2 and cavities in IRAS4C, can be explained by means of different evolutionary stages. 
The outflow from IRAS4C having a wider opening angle is likely to be more evolved than that of IRAS 4A2 \cite[][]{machida2013, arce2006}.

\section{CONCLUSIONS} \label{sec:conclusions}
This paper presents the reduced image from ALMA archival data of Class 0 source NGC1333 IRAS 4A in Perseus molecular cloud. The resolutions of these images are 0.2$\arcsec$ (60 au) for the 1.3 mm continuum, $^{12}$CO $(J=2-1)$, SO $(J_N=6_5-5_4)$ lines, and 0.6$\arcsec$ (180 au) for the 1.17 mm continuum, the SO $(J_N=7_6-6_5)$, CCH ($J_{N,F}=7/2_{3,4}-5/2_{2,3}$) lines. The analysis of these images provides the following findings of kinematics in IRAS 4A. 

\begin{enumerate}
\item
The high resolution of CO $(J=2-1)$ and SO $(J_N=6_5-5_4)$ line observations revealed that IRAS 4A1 and IRAS 4A2 are driving independent outflows. The northern outflow lobes are overlapped each other at roughly $\sim 3\arcsec$ north the IRAS 4A1 and IRAS 4A2 while southern outflows propagate in different directions. 

\item
In the IRAS 4A1 and IRAS 4A2 outflows, the column density and the rotation temperature estimated from two transitions of SO emission increase toward the vicinity of IRAS 4A1 and IRAS 4A2 continuum peaks.

\item
IRAS 4A2 drives S-shaped outflow from the edges of a flattened rotational envelope centered at IRAS 4A2's continuum peak. The outflow shows different degrees of asymmetry at different velocity ranges. Additionally, CO observation reveals extremely high velocity components at the base of the IRAS 4A2 outflow. These features may be explained by magnetic field misalignment model, in which the initial core angular momentum vector is misaligned with the magnetic field, or the precession caused by a close unresolved binary system, with a separation larger than $\sim$12 au.

\item
The flattened envelope around IRAS 4A2 has an opposite velocity gradient with the circumbinary envelope. This could possibly be explained by Hall effect: The Hall induced magnetic field may flip the angular momentum vector between the flattened envelope and the circumstellar envelope.

\item
The CCH ($J_{N,F}=7/2_{3,4}-5/2_{2,3}$) emission shows two pairs of blobs attaching to the bottom of shell like feature, and the morphology is significantly different from emission of other molecular lines. We suggest that CCH emission is enhanced in the outflow cavity walls. 
\end{enumerate}

\acknowledgments
This paper makes use of the following ALMA data: ADS/JAO.ALMA2013.1.01102.S, ADS/JAO.A-LMA2013.1.00031.S, and ADS/JAO.ALMA2017.1.00053.S. ALMA is a partnership of ESO (representing its member states), NSF (USA) and NINS (Japan), together with NRC (Canada), MOST and ASIAA (Taiwan), and KASI (Republic of Korea), in cooperation with the Republic of Chile. The Joint ALMA Observatory is operated by ESO, AUI/NRAO and NAOJ.

N.H. acknowledges a grant from the Ministry of Science and Technology (MoST) of
Taiwan (MoST 108-2112-M-001-017-, MoST 109-2112-M-001-023-).

\vspace{5mm}
\facilities{ALMA}

\software{CASA \citep{mcmu07}, MIRIAD \citep{saul95}, Radmc3D \cite[][]{radmc3d2012}, vis\_sample (For further references see \href{https://github.com/AstroChem/vis_sample.git}{vis sample}), Astropy \citep{astropy:2013, astropy:2018}, Aplpy \citep{robitaille2012}} 
\newpage

\appendix
\section{Channel Maps}
\label{sec:app_ch}

\begin{figure}[ht!]
\epsscale{2.0}
\hspace*{-6.0cm}
\plotone{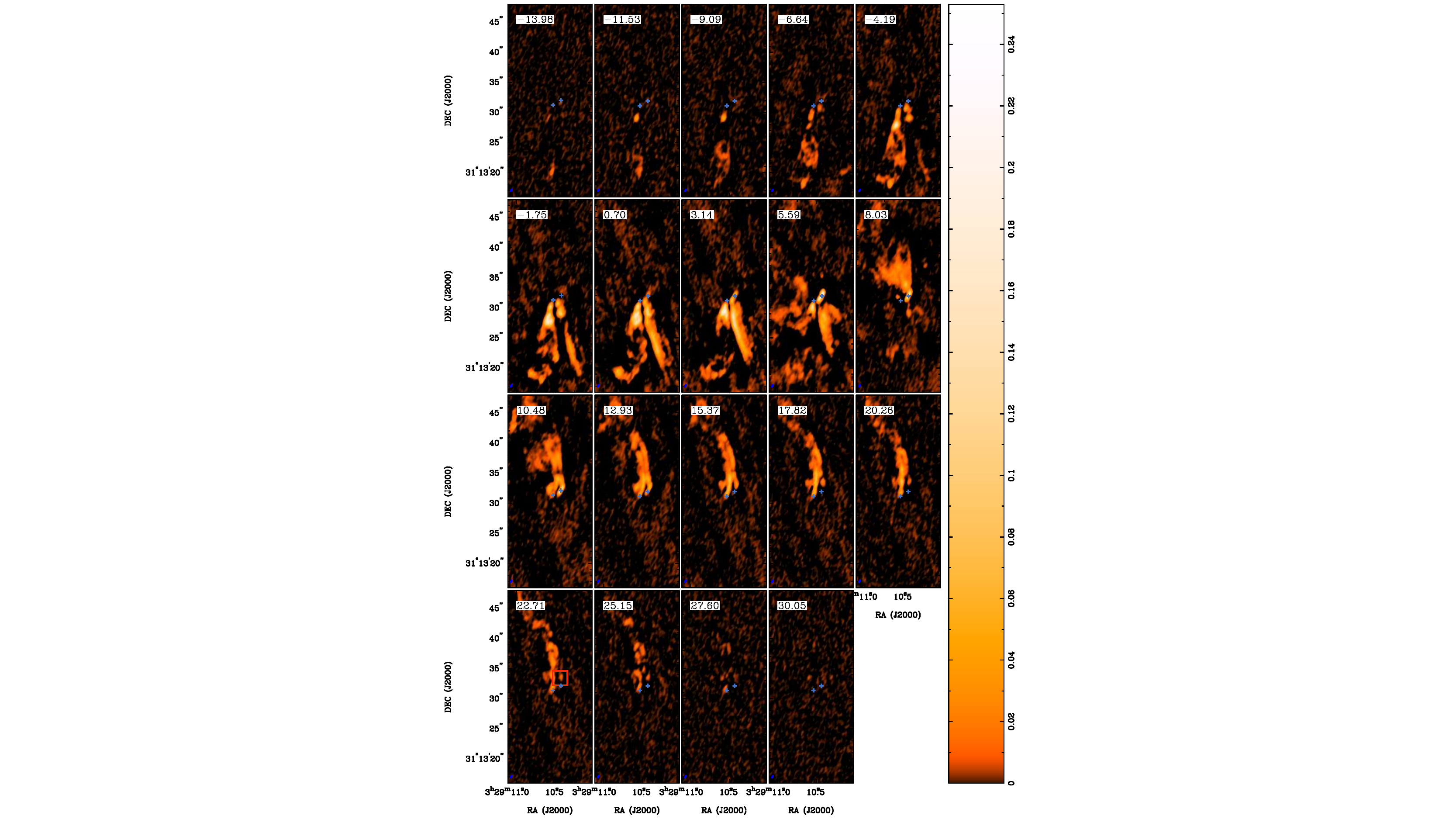}
\caption{Channel maps of the SO $J_N=7_6-6_5$ emission line. The velocity resolution is $2.44\ \kms$. The blue ellipse in each panel is the ALMA synthesized beam, $0\farcs 65\times 0\farcs 35,\ {\rm P.A.}=-28^{\circ}$, respectively. The red box is the viewing range of Figure \ref{fig:so3ch}. 
\label{fig:so7ch}}
\end{figure}
\newpage

\begin{figure}[ht!]
\epsscale{1.2}
\plotone{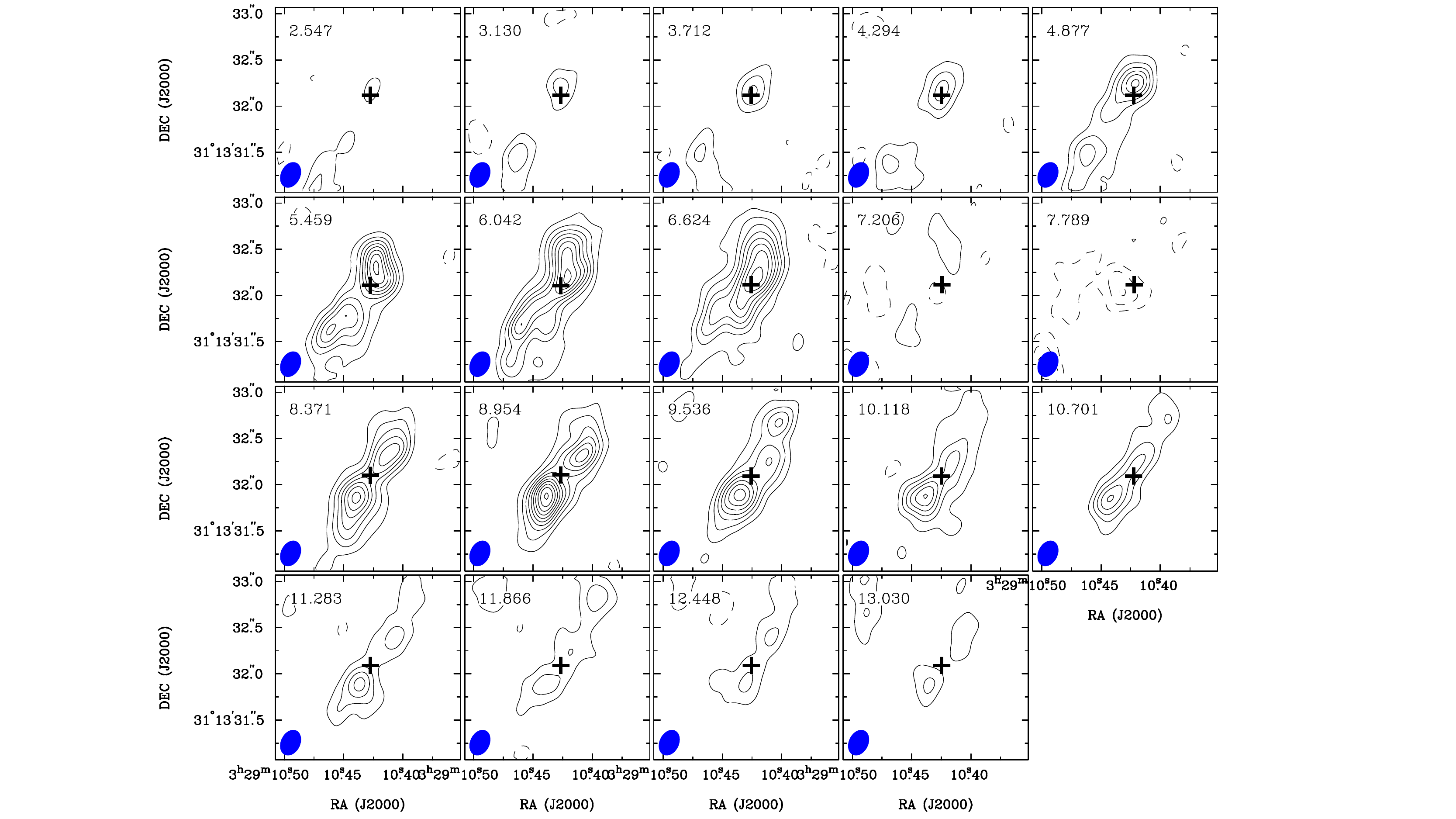}
\caption{Channel maps of the SO $J_N=6_5-5_4$ emission line (centered at IRAS 4A2). The velocity resolution is $0.58\ \kms$. Contour levels are from $3\sigma$ in steps of $3\sigma$, where $1\sigma$ corresponds to $3\ \mJB$. The blue ellipse in each panel is the ALMA synthesized beam, $0\farcs 29\times 0\farcs 21,\ {\rm P.A.}=-26^{\circ}$, respectively.
\label{fig:so3ch}}
\end{figure}
\newpage

\begin{figure*}[ht!]
\epsscale{2.0}
\hspace*{-6.5cm}
\plotone{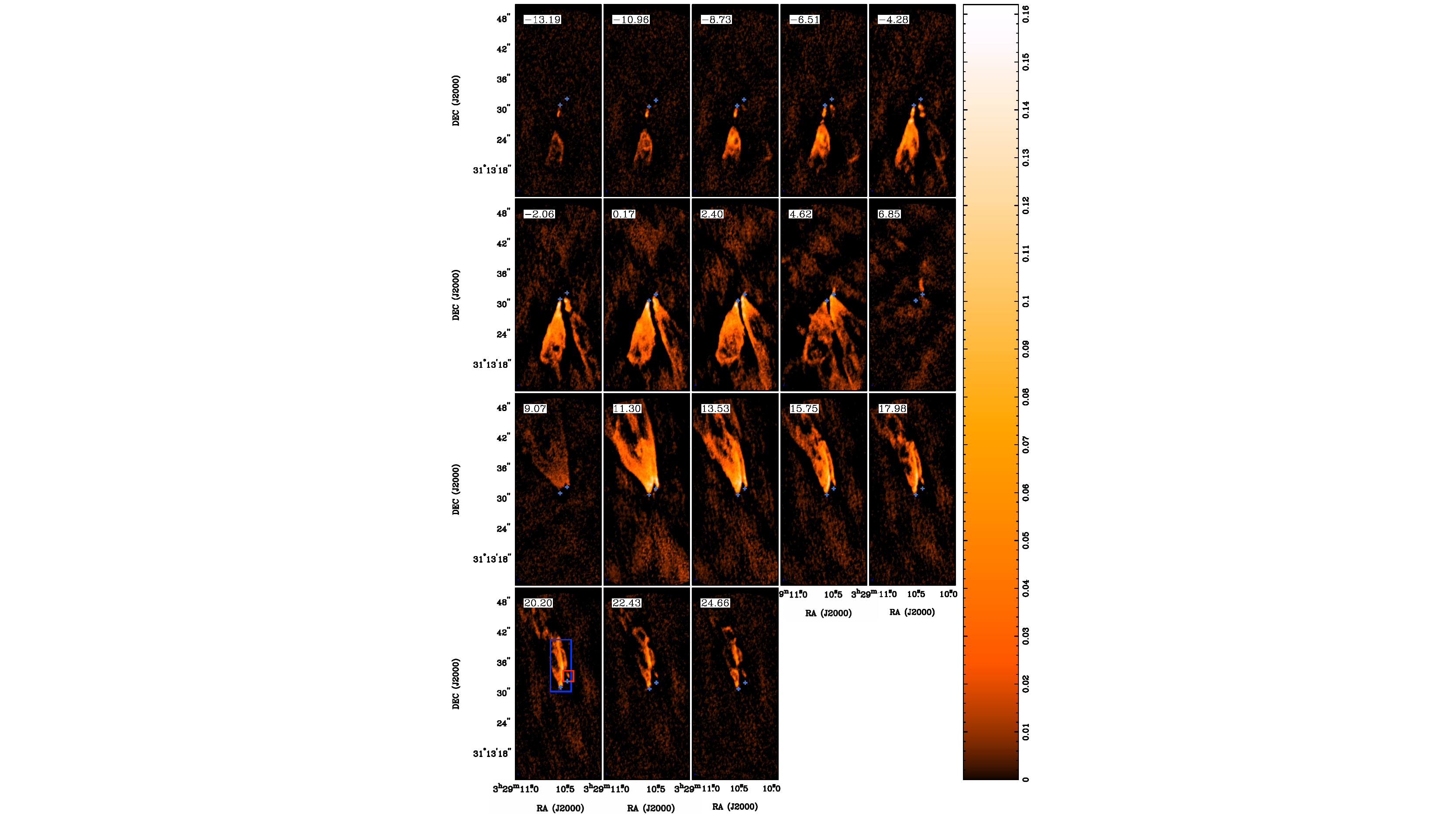}
\caption{Channel maps of the CO $J=2-1$ emission line (low velocity to high velocity). The velocity resolution is $2.2\ \kms$. The blue ellipse in each panel is the ALMA synthesized beam, $0\farcs 29\times 0\farcs 21,\ {\rm P.A.}=-26^{\circ}$, respectively. The blue box is the viewing range of Figure \ref{fig:cohvch} and the red box is the viewing range of Figure \ref{fig:coa2ch}.
\label{fig:convch}}
\end{figure*}
\newpage

\begin{figure*}[ht!]
\epsscale{1.9}
\hspace*{-5.5cm}
\plotone{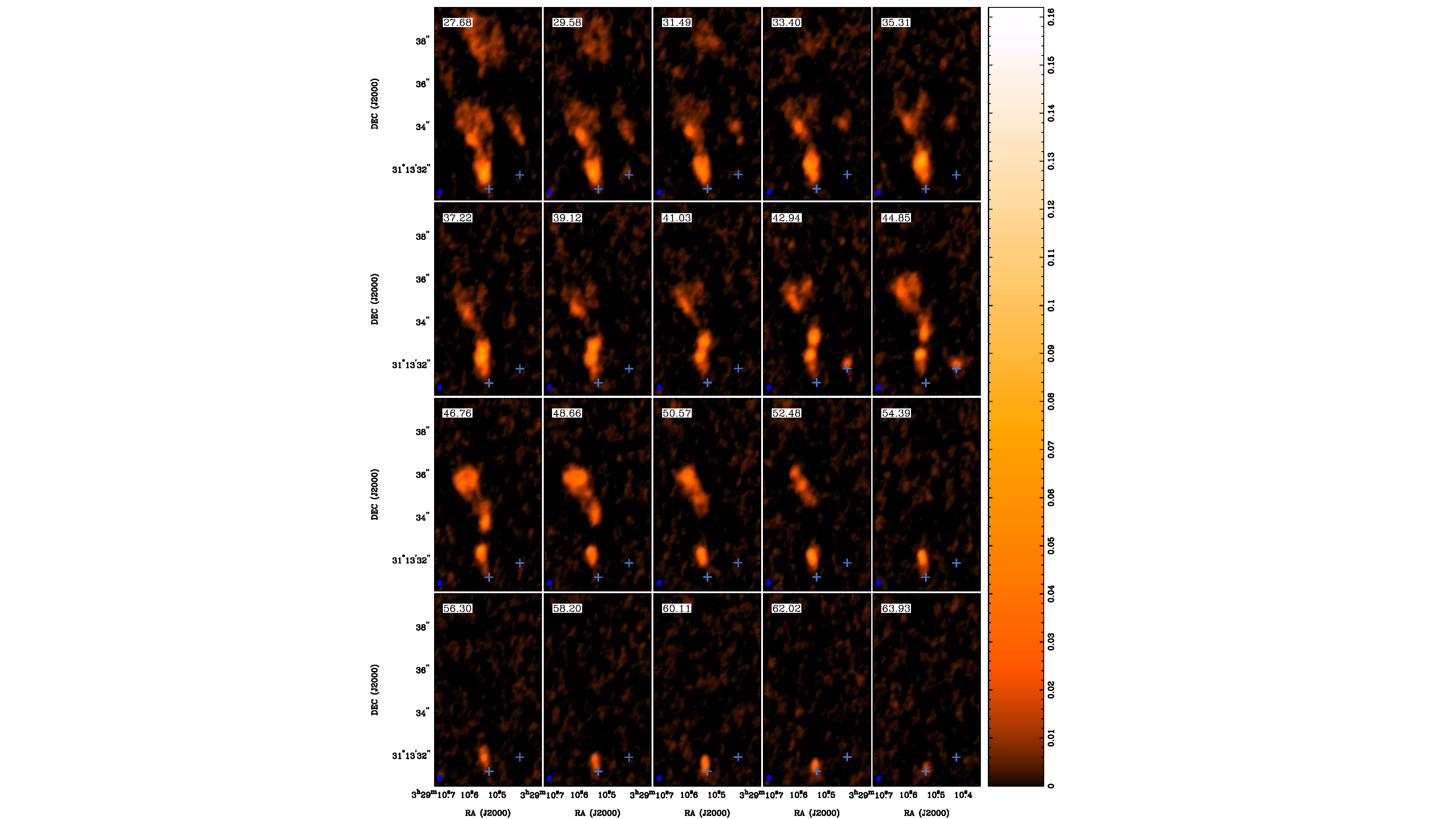}
\caption{Channel maps of the CO $J=2-1$ emission line (very high velocity). The velocity resolution is $1.9\ \kms$. The blue ellipse in each panel is the ALMA synthesized beam, $0\farcs 29\times 0\farcs 21,\ {\rm P.A.}=-26^{\circ}$, respectively.
\label{fig:cohvch}}
\end{figure*}
\newpage

\begin{figure}[ht!]
\epsscale{1.0}
\plotone{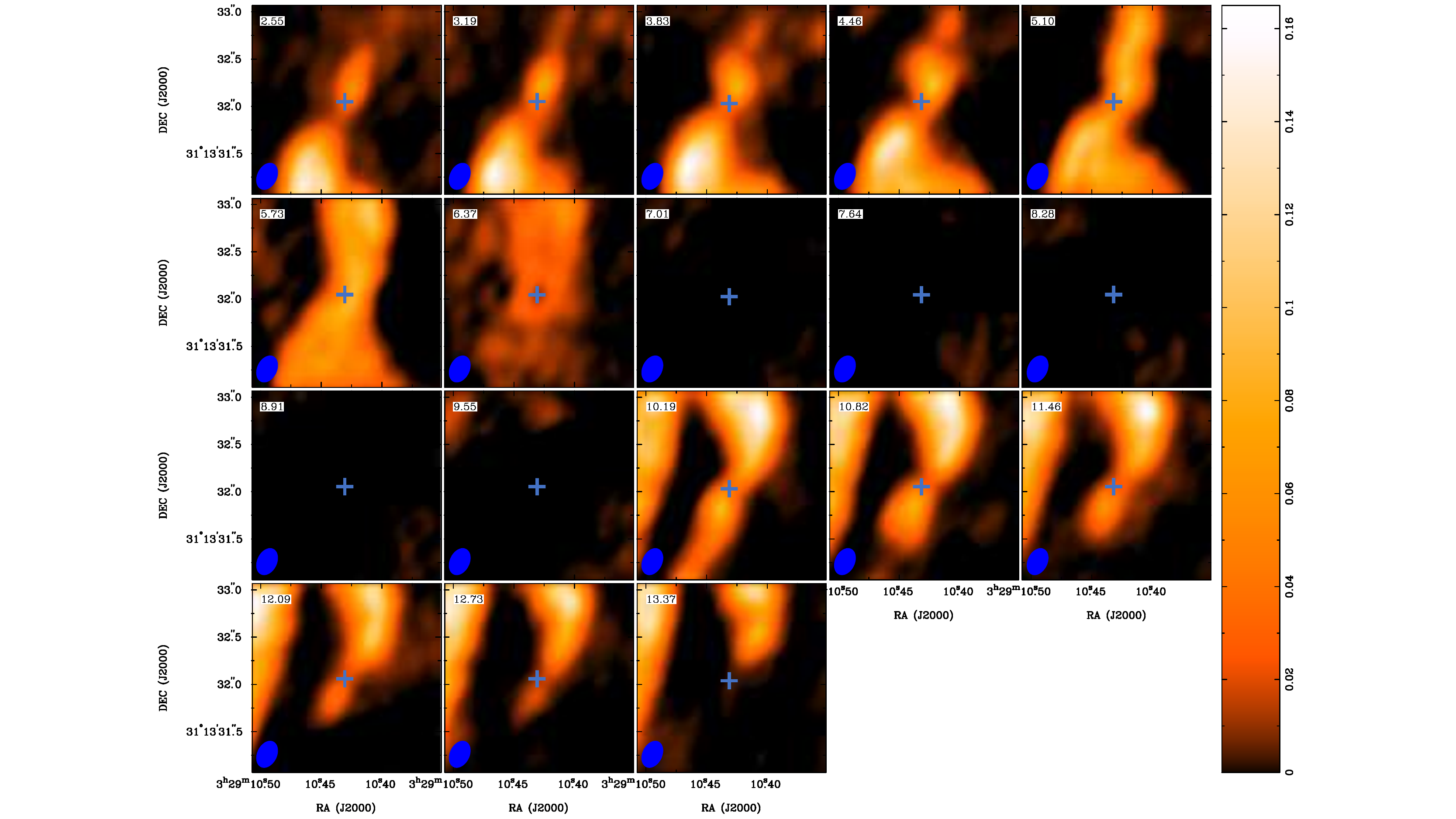}
\caption{Channel maps of the CO $J=2-1$ emission line (centered at IRAS 4A2). The velocity resolution is $0.64\ \kms$. The blue ellipse in each panel is the ALMA synthesized beam, $0\farcs 29\times 0\farcs 21,\ {\rm P.A.}=-26^{\circ}$, respectively.
\label{fig:coa2ch}}
\end{figure}
\newpage

\begin{figure}[ht!]
\epsscale{1.5}
\hspace*{-0.2cm}
\plotone{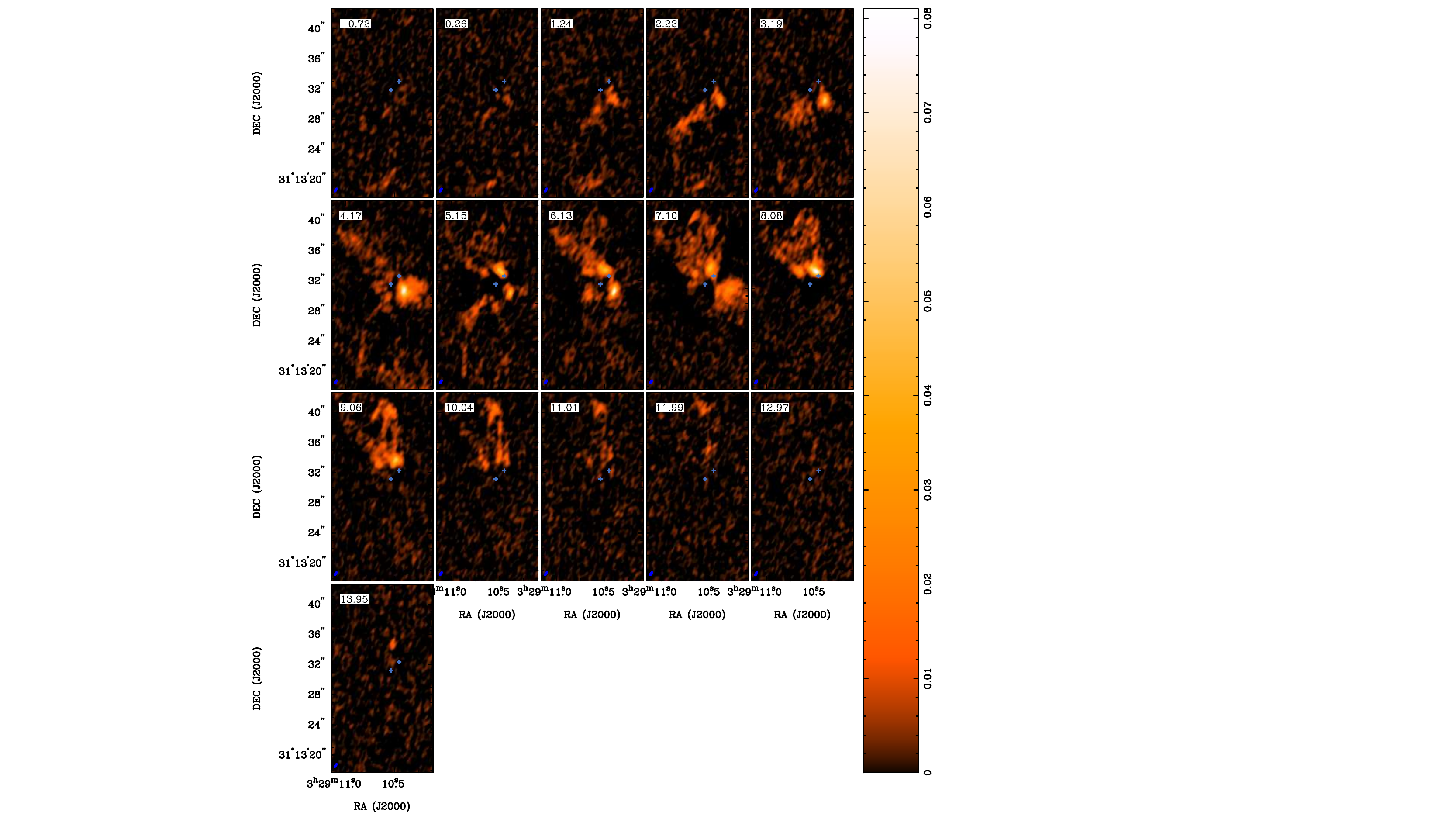}
\caption{Channel maps of the CCH $N=3-2$ emission line. The velocity resolution is $1.0\ \kms$. The blue ellipse in each panel is the ALMA synthesized beam, $0\farcs 65\times 0\farcs 35,\ {\rm P.A.}=-28^{\circ}$, respectively.
\label{fig:cchch}}
\end{figure}
\newpage

\clearpage
\section{SO ($J_N=6_5-5_4$) Moment Maps}
\label{sec:app_sol}

\begin{figure}[ht!]
\epsscale{0.8}
\plotone{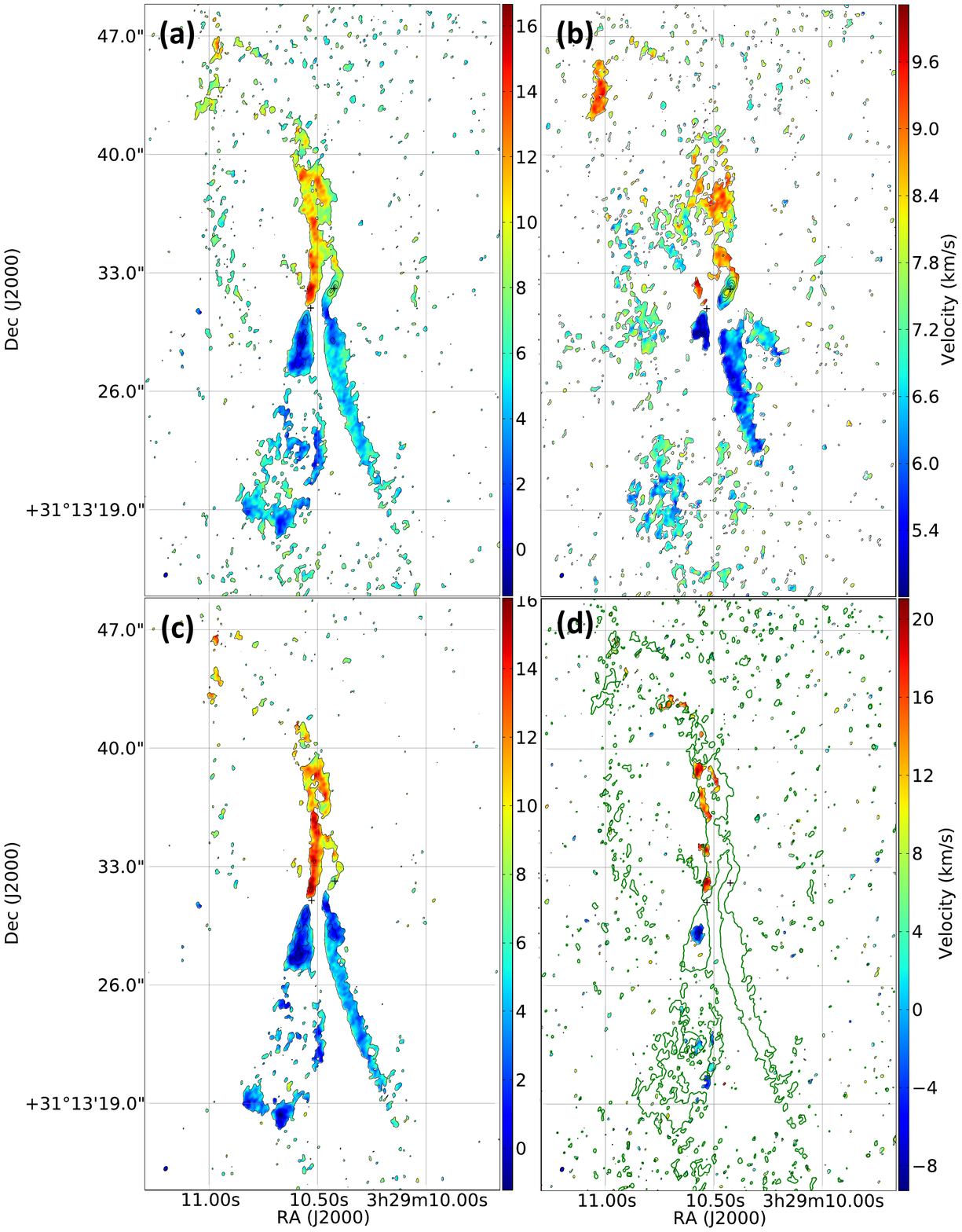}
\caption{Same as Figure \ref{fig:so}, but for SO ($J_N=6_5-5_4$) transition. The contour levels are $1,2,3,4,\dots \times 3\sigma$. 1 $\sigma$ corresponds to (a) 46.5, (b) 8.7, (c) 26.9, and (d) $34.1\ \mJB~\kms$. 
\label{fig:app_sol}}
\end{figure}


\newpage

\clearpage
\begin{thebibliography}{}


\bibitem[Akeson, \& Carlstrom(1997)]{akeson1997} Akeson, R.~L., \& Carlstrom, J.~E.\ 1997, \apj, 491, 254
\bibitem[Arce \& Sargent(2006)]{arce2006} Arce, H.~G., \& Sargent, A.~I.\ 2006, \apj, 646, 1070
\bibitem[Aso et al.(2018)]{aso2018} Aso, Y., Hirano, N., Aikawa, Y., et al.\ 2018, \apj, 863, 19
\bibitem[Astropy Collaboration et al.(2013)]{astropy:2013} Astropy Collaboration, Robitaille, T.~P., Tollerud, E.~J., et al.\ 2013, \aap, 558, A33
\bibitem[Astropy Collaboration et al.(2018)]{astropy:2018} Astropy Collaboration, Price-Whelan, A.~M., Sip{\H{o}}cz, B.~M., et al.\ 2018, \aj, 156, 123
\bibitem[Attard et al.(2009)]{attard2009} Attard, M., Houde, M., Novak, G., et al.\ 2009, \apj, 702, 1584
\bibitem[Banerjee, \& Pudritz(2006)]{Banerjee2006} Banerjee, R., \& Pudritz, R.~E.\ 2006, \apj, 641, 949
\bibitem[Bate et al.(2000)]{bate2000} Bate, M.~R., Bonnell, I.~A., Clarke, C.~J., et al.\ 2000, \mnras, 317, 773. doi:10.1046/j.1365-8711.2000.03648.x
\bibitem[Ching et al.(2016)]{Ching2016} Ching, T.-C., Lai, S.-P., Zhang, Q., et al.\ 2016, \apj, 819, 159
\bibitem[Choi et al.(2010)]{Choi2010} Choi, M., Tatematsu, K., \& Kang, M.\ 2010, \apjl, 723, L34
\bibitem[Choi et al.(2011)]{Choi2011} Choi, M., Kang, M., Tatematsu, K., et al.\ 2011, \pasj, 63, 1281
\bibitem[Choi(2001)]{Choi2001} Choi, M.\ 2001, \apj, 553, 219
\bibitem[Choi(2005)]{choi2005} Choi, M.\ 2005, \apj, 630, 976
\bibitem[Choi et al.(2006)]{choi2006} Choi, M., Hodapp, K.~W., Hayashi, M., et al.\ 2006, \apj, 646, 1050. doi:10.1086/505037
\bibitem[Ciardi \& Hennebelle(2010)]{ciardi2010} Ciardi, A., \& Hennebelle, P.\ 2010, \mnras, 409, L39
\bibitem[Doi et al.(2020)]{doi2020} Doi, Y., Hasegawa, T., Furuya, R.~S., et al.\ 2020, \apj, 899, 28. doi:10.3847/1538-4357/aba1e2
\bibitem[Dullemond et al.(2012)]{radmc3d2012} Dullemond, C.~P., Juhasz, A., Pohl, A., et al.\ 2012, RADMC-3D: A multi-purpose radiative transfer tool, ascl:1202.015
\bibitem[Goldsmith \& Langer(1999)]{goldsmith1999} Goldsmith, P.~F. \& Langer, W.~D.\ 1999, \apj, 517, 209. doi:10.1086/307195
\bibitem[Goodwin et al. a (2004)]{goodwin2004I} Goodwin, S.~P., Whitworth, A.~P., \& Ward-Thompson, D.\ 2004, \aap, 414, 633
\bibitem[Goodwin et al. b (2004)]{goodwin2004II} Goodwin, S.~P., Whitworth, A.~P., \& Ward-Thompson, D.\ 2004, \aap, 423, 169
\bibitem[Goodwin et al.(2006)]{goodwin2006} Goodwin, S.~P., Whitworth, A.~P., \& Ward-Thompson, D.\ 2006, \aap, 452, 487
\bibitem[Gueth et al.(1996)]{gueth1996} Gueth, F., Guilloteau, S., \& Bachiller, R.\ 1996, \aap, 307, 891
\bibitem[Harsono et al.(2014)]{harsono2014} Harsono, D., J{\o}rgensen, J.~K., van Dishoeck, E.~F., et al.\ 2014, \aap, 562, A77
\bibitem[Hildebrand(1983)]{Hildebrand1983} Hildebrand, R.~H.\ 1983, \qjras, 24, 267
\bibitem[Hirano et al.(2010)]{hirano2010} Hirano, N., Ho, P.~P.~T., Liu, S.-Y., et al.\ 2010, \apj, 717, 58. doi:10.1088/0004-637X/717/1/58
\bibitem[Hirano \& Machida(2019)]{Hirano2019} Hirano, S., \& Machida, M.~N.\ 2019, \mnras, 485, 4667
\bibitem[Hull et al.(2013)]{Hull2013} Hull, C.~L.~H., Plambeck, R.~L., Bolatto, A.~D., et al.\ 2013, \apj, 768, 159
\bibitem[Hull et al.(2014)]{hull2014} Hull, C.~L.~H., Plambeck, R.~L., Kwon, W., et al.\ 2014, VizieR Online Data Catalog, J/ApJS/213/13
\bibitem[Joos et al.(2012)]{Joos2012} Joos, M., Hennebelle, P., \& Ciardi, A.\ 2012, \aap, 543, A128
\bibitem[Joos et al.(2013)]{joos2013} Joos, M., Hennebelle, P., Ciardi, A., et al.\ 2013, \aap, 554, A17
\bibitem[Kramer et al.(1998)]{kramer1998} Kramer, C., Alves, J., Lada, C., et al.\ 1998, \aap, 329, L33
\bibitem[Kwon et al.(2015)]{kwon2015} Kwon, W., Fern{\'a}ndez-L{\'o}pez, M., Stephens, I.~W., et al.\ 2015, \apj, 814, 43. doi:10.1088/0004-637X/814/1/43
\bibitem[Lee et al.(2000)]{lee2000} Lee, C.-F., Mundy, L.~G., Reipurth, B., et al.\ 2000, \apj, 542, 925. doi:10.1086/317056
\bibitem[Lee et al.(2010)]{Lee2010_1} Lee, C.-F., Hasegawa, T.~I., Hirano, N., et al.\ 2010, \apj, 713, 731
\bibitem[Lee et al.(2010)]{lee2010} Lee, C.-F., Hasegawa, T.~I., Hirano, N., et al.\ 2010, \apj, 713, 731
\bibitem[Lee(2010)]{Lee2010} Lee, C.-F.\ 2010, \apj, 725, 712
\bibitem[Li et al.(2013)]{Li2013} Li, Z.-Y., Krasnopolsky, R., \& Shang, H.\ 2013, \apj, 774, 82
\bibitem[Li et al.(2017)]{Li2017} Li, J.~I.-H., Liu, H.~B., Hasegawa, Y., et al.\ 2017, \apj, 840, 72
\bibitem[Lindberg et al.(2014)]{Lindberg2014} Lindberg, J.~E., J{\o}rgensen, J.~K., Brinch, C., et al.\ 2014, \aap, 566, A74
\bibitem[Looney et al.(2000)]{Looney2000} Looney, L.~W., Mundy, L.~G., \& Welch, W.~J.\ 2000, \apj, 529, 477
\bibitem[L{\'o}pez-Sepulcre et al.(2017)]{Sepulcre2017} L{\'o}pez-Sepulcre, A., Sakai, N., Neri, R., et al.\ 2017, \aap, 606, A121
\bibitem[Machida \& Basu(2019)]{Machida2019} Machida, M.~N. \& Basu, S.\ 2019, \apj, 876, 149
\bibitem[Machida et al.(2008)]{machida2008} Machida, M.~N., Inutsuka, S.-. ichiro ., \& Matsumoto, T.\ 2008, \apj, 676, 1088. doi:10.1086/528364
\bibitem[Machida et al.(2011)]{machida11} Machida, M.~N., Inutsuka, S.-I., \& Matsumoto, T.\ 2011, \pasj, 63, 555
\bibitem[Machida et al.(2020)]{machida2020} Machida, M.~N., Hirano, S., \& Kitta, H.\ 2020, \mnras, 491, 2180. doi:10.1093/mnras/stz3159
\bibitem[Machida \& Hosokawa(2013)]{machida2013} Machida, M.~N., \& Hosokawa, T.\ 2013, \mnras, 431, 1719
\bibitem[Machida \& Hosokawa(2013)]{machida2013ii} Machida, M. \& Hosokawa, T.\ 2013, Protostars and Planets VI Posters
\bibitem[Machida \& Matsumoto(2012)]{machida2012} Machida, M.~N. \& Matsumoto, T.\ 2012, \mnras, 421, 588. doi:10.1111/j.1365-2966.2011.20336.x
\bibitem[Matsumoto \& Tomisaka(2004)]{matsumoto2004} Matsumoto, T., \& Tomisaka, K.\ 2004, \apj, 616, 266
\bibitem[Matsumoto et al.(2017)]{matsumoto2017} Matsumoto, T., Machida, M.~N., \& Inutsuka, S.-. ichiro .\ 2017, \apj, 839, 69
\bibitem[Matthews et al.(2009)]{Matthews2009} Matthews, B.~C., McPhee, C.~A., Fissel, L.~M., et al.\ 2009, \apjs, 182, 143
\bibitem[Maury et al.(2019)]{maury2019} Maury, A.~J., Andr{\'e}, P., Testi, L., et al.\ 2019, \aap, 621, A76. doi:10.1051/0004-6361/201833537
\bibitem[McMullin et al.(2007)]{mcmu07} McMullin, J.~P., Waters, B., Schiebel, D., Young, W., \& Golap, K.\ 2007, Astronomical Data Analysis Software and Systems XVI, 376, 127
\bibitem[Myers et al.(2013)]{myers2013} Myers, A.~T., McKee, C.~F., Cunningham, A.~J., et al.\ 2013, \apj, 766, 97
\bibitem[M{\'e}nard, \& Duch{\^e}ne(2004)]{Menard2004} M{\'e}nard, F., \& Duch{\^e}ne, G.\ 2004, \aap, 425, 973
\bibitem[Nakamura et al.(1995)]{nakamura1995} Nakamura, F., Hanawa, T., \& Nakano, T.\ 1995, \apj, 444, 770. doi:10.1086/175650
\bibitem[Ohashi et al.(2014)]{Ohashi2014} Ohashi, N., Saigo, K., Aso, Y., et al.\ 2014, \apj, 796, 131
\bibitem[Ohashi et al.(2014)]{ohashi2014} Ohashi, N., Saigo, K., Aso, Y., et al.\ 2014, \apj, 796, 131
\bibitem[Ossenkopf \& Henning(1994)]{ossenkopf1994} Ossenkopf, V. \& Henning, T.\ 1994, \aap, 291, 943
\bibitem[Park, \& Choi(2007)]{choi07} Park, G., \& Choi, M.\ 2007, \apjl, 664, L99
\bibitem[Pety et al.(2005)]{pety2005} Pety, J., Teyssier, D., Foss{\'e}, D., et al.\ 2005, \aap, 435, 885
\bibitem[Raga et al.(2009)]{raga2009} Raga, A.~C., Esquivel, A., Vel{\'a}zquez, P.~F., et al.\ 2009, \apjl, 707, L6. doi:10.1088/0004-637X/707/1/L6
\bibitem[Reipurth et al.(2002)]{Reipurth2002} Reipurth, B., Rodr{\'\i}guez, L.~F., Anglada, G., et al.\ 2002, \aj, 124, 1045
\bibitem[Robitaille(2011)]{hyperion2011} Robitaille, T.~P.\ 2011, \aap, 536, A79
\bibitem[Robitaille \& Bressert(2012)]{robitaille2012} Robitaille, T. \& Bressert, E.\ 2012, Astrophysics Source Code Library
\bibitem[Sahu et al.(2019)]{Sahu2019} Sahu, D., Liu, S.-Y., Su, Y.-N., et al.\ 2019, \apj, 872, 196
\bibitem[Santangelo et al.(2015)]{Santangelo2015} Santangelo, G., Codella, C., Cabrit, S., et al.\ 2015, \aap, 584, A126
\bibitem[Sault et al.(1995)]{saul95} Sault, R.~J., Teuben, P.~J., \& Wright, M.~C.~H.\ 1995, Astronomical Data Analysis Software and Systems IV, 77, 433 
\bibitem[Seifried et al.(2013)]{seifried2013} Seifried, D., Banerjee, R., Pudritz, R.~E., et al.\ 2013, \mnras, 432, 3320
\bibitem[Shepherd \& Watson(2002)]{shepherd2002} Shepherd, D.~S. \& Watson, A.~M.\ 2002, \apj, 566, 966. doi:10.1086/338138
\bibitem[Stephens et al.(2018)]{stephens2018} Stephens, I.~W., Dunham, M.~M., Myers, P.~C., et al.\ 2018, \apjs, 237, 22
\bibitem[Su et al.(2019)]{su2019} Su, Y.-N., Liu, S.-Y., Li, Z.-Y., et al.\ 2019, arXiv e-prints, arXiv:1909.12443
\bibitem[Takakuwa et al.(2018)]{takakuwa2018} Takakuwa, S., Tsukamoto, Y., Saigo, K., et al.\ 2018, \apj, 865, 51
\bibitem[Taquet et al.(2020)]{taquet2020} Taquet, V., Codella, C., De Simone, M., et al.\ 2020, \aap, 637, A63. doi:10.1051/0004-6361/201937072
\bibitem[Tobin et al.(2016)]{tobin2016} Tobin, J.~J., Looney, L., Li, Z.-Y., et al.\ 2016, American Astronomical Society Meeting Abstracts \#227
\bibitem[Tobin et al.(2018)]{tobin2018} Tobin, J.~J., Looney, L.~W., Li, Z.-Y., et al.\ 2018, VizieR Online Data Catalog, J/ApJ/818/73
\bibitem[Tobin et al.(2018)]{tobin2018ii} Tobin, J.~J., Looney, L.~W., Li, Z.-Y., et al.\ 2018, \apj, 867, 43. doi:10.3847/1538-4357/aae1f7
\bibitem[Tomisaka(1996)]{tomisaka1996} Tomisaka, K.\ 1996, \pasj, 48, L97. doi:10.1093/pasj/48.5.L97
\bibitem[Tsukamoto et al.(2017)]{tsukamoto2017} Tsukamoto, Y., Okuzumi, S., Iwasaki, K., et al.\ 2017, \pasj, 69, 95
\bibitem[Wilson \& Rood(1994)]{wilson1994} Wilson, T.~L. \& Rood, R.\ 1994, \araa, 32, 191. doi:10.1146/annurev.aa.32.090194.001203
\bibitem[Y{\i}ld{\i}z et al.(2012)]{yildiz2012} Y{\i}ld{\i}z, U.~A., Kristensen, L.~E., van Dishoeck, E.~F., et al.\ 2012, \aap, 542, A86
\bibitem[Zhang et al.(2018)]{zhang2018} Zhang, Y., Higuchi, A.~E., Sakai, N., et al.\ 2018, \apj, 864, 76
\bibitem[Zucker et al.(2018)]{Zucker2018} Zucker, C., Schlafly, E.~F., Speagle, J.~S., et al.\ 2018, \apj, 869, 83


\end{thebibliography}
\end{document}